\documentclass{article}

\usepackage{arxiv}

\usepackage[utf8]{inputenc} % allow utf-8 input
\usepackage[T1]{fontenc}    % use 8-bit T1 fonts
\usepackage{hyperref}       % hyperlinks
\usepackage{url}            % simple URL typesetting
\usepackage{booktabs}       % professional-quality tables
\usepackage{amsfonts}       % blackboard math symbols
\usepackage{nicefrac}       % compact symbols for 1/2, etc.
\usepackage{float}
\usepackage{microtype}      % microtypography
\usepackage{lipsum}
\usepackage{fancyhdr}
\usepackage{amsmath}
\usepackage{pifont}
\usepackage{booktabs}
\usepackage{dcolumn} 
\usepackage{multirow}
\usepackage{graphicx}       % graphics
\graphicspath{{imgs/}}     % organize your images and other figures under media/ folder

\title{The Importance of (Exponentially More) \\ Computing Power
}

\author
{Neil C. Thompson$^{1\ast}$, Shuning Ge$^{2}$, Gabriel F. Manso$^{3}$\\
\\
\normalsize{$^{1}$MIT Computer Science and A.I. Lab,}\\ \normalsize{MIT Initiative on the Digital Economy, Cambridge, MA USA}\\
\normalsize{$^{2}$MIT, Cambridge MA, USA}\\
\normalsize{$^{3}$FGA,  University of Brasilia, Brasilia, Brazil} \\
\\
\normalsize{$^\ast$To whom correspondence should be addressed; E-mail:  neil\_t@mit.edu.}}

\begin{document}
\maketitle

\begin{abstract}
    Denizens of Silicon Valley have called Moore’s Law “the most important graph in human history,” and economists have found that Moore’s Law-powered I.T. revolution has been one of the most important sources of national productivity growth. But data substantiating these claims tend to either be abstracted --- for example by examining spending on I.T., rather than I.T. itself --- or anecdotal. In this paper, we assemble direct quantitative evidence of the impact that computing power has had on five domains: two computing bellwethers (Chess and Go), and three economically important applications (weather prediction, protein folding, and oil exploration).  Computing power explains 49\%-94\% of the performance improvements in these domains. But whereas economic theory typically assumes a power law relationship between inputs and outputs, we find that an \textit{exponential} increase in computing power is needed to get \textit{linear} improvements in these outcomes. This helps clarify why the exponential growth of computing power from Moore's Law has been so important for progress, and why performance improvements across many domains are becoming economically tenuous as Moore’s Law breaks down.
\end{abstract}

% keywords can be removed

\section{Introduction}\label{intro}
The key question at the heart of this paper is how more powerful computers are improving outcomes across society. We analyze this question by examining the growing use of computing power increases across five key application areas and then estimating those production functions. We find that systems designers have paid handsomely to grow computing power at a rate faster than the underlying hardware progress, but that these investments have paid off in the form of more capable, better-performing systems. Collectively, our results highlight the important role that exponential improvement in computing power have had for generating progress across diverse applications.

Production functions, be they macroeconomic models of the economy or microeconomic models of individual agents or firms, attempt to model and parameterize how changes in key inputs produce changes in output. Traditional models, for example \cite{romer_1989, romer_1990}, focused on two inputs: labor and capital, which are mutually complementary, but independently face decreasing marginal returns. But not all capital is the same, and researchers have argued that I.T. capital should be split from other forms of capital \cite{brynjolfsson_1996, brynjolfsson_1997}, consistent with the popular view of data-driven business, where I.T. capital is more, not less, important to businesses with large amounts of other capital (e.g. computer-driven planning is more important for FedEx because it has so many trucks). Some modern growth models now explicitly recognize this elevated role for I.T. in production functions and its complementarity with other forms of capital \cite{acemoglu_2018, acemoglu_2018_race, cheng_2007}. 

At the macro-level, the elevation of I.T. capital as primary input in production functions has been validated by studies showing that I.T. has contributed more than one-third of all improvement in productivity since 1974 \cite{byrne_2013}, as well as earlier studies showing that IT contributed nearly all the improvements in total factor productivity from in the late 1990s \cite{jorgenson_2002, oliner_2002}. The contributions of I.T. also extend beyond the economics, to include other measures of well-being \cite{ganju_2016}.  

Studies at the micro-level have found similarly strong effects on the importance of I.T. on firm productivity \cite{banker1991reuse, gerow2014looking, sabherwal2015information, foster1984management, pinsonneault1998information}. For example Brynjolfsson and Hitt \cite{brynjolfsson_1996} showed I.T. capital, as distinct from other types of capital, contributed significantly to marginal firm output and Thompson showed that a substantial fraction of growth in firm total factor productivity after 2005 could be explained by better abilities to harness modern computer chips \cite{Thompson2017}. Most studies, however, take an abstracted view of computing capital, measuring inputs in dollars spent rather than in the productive capacity of what was bought. As Devaraj and Kholi \cite{devaraj2003performance} pointed out, this is inferior to measuring I.T. in the form of actual usage. And while this same critique could be made for almost all input factors to production functions, the distinction between spend and productive capacity is more strongly divergent for computing power because of exponential increases in computing power per dollar\footnote{Official statistics do attempt to address this issue via deflators, for example based on the SPECInt computing benchmark, but these make the strong implicit assumption that computing power has the same value in different domains \cite{byrne2018fast} which we later show is problematic.}. This mismatch is evidenced, for example, by the finding that \$1 in computer hardware spend is associated with \$10 of increased firm value \cite{saunders2015valuing}).

In this paper, we also study the micro-foundations of how I.T. capital improves performance, but we do it in the natural units of computer processors: computing power, as measured by the number of operations that the processor can perform. This is very much in the spirit of Thompson, Greenewald, Lee, and Manso \cite{thompson2020computational}, which shows that many of the most-important improvements in machine learning are heavily dependent on improvements in computing power. In this paper, we extend this type of analysis to other areas, and show that a much broader conclusion can be reached: progress across many areas of computing is dependent on exponential increases in computing power. To reach this conclusion, we gather detailed records on the performance and usage of computing power across five domains.

Chess and Go are important bellwethers for computing performance because both were traditionally viewed as areas of human expertise. Therefore, progress against human acumen could be used to track the development of programs that could play these games. For example, chess master David Levy said “Until 1977, there seemed to be no point in my playing a formal challenge match against any chess program because none of them were good enough, but when [the program] CHESS 4.5 began doing well… it was time for me to defend the human race against the coming invasion.” \cite{levy1982all}. Of the two games, Go is much harder. In 1997, astrophysicist Piet Hut, from the Institute for Advanced Study in New Jersey, told the New York Times (in retrospect incorrectly) that “It may be a hundred years before a computer beats humans at Go — maybe even longer.'' \cite{muoio2016ai}. Because these two games are such bellwethers, they have attracted substantial attention from computer scientists which has led to a broad exploration of computational algorithms, hardware and software approaches for playing. As such, they present a promising way to study the importance of computing power for progress.

We also consider the impact of computing power in three economically-important settings: weather prediction, protein folding, and oil exploration. Weather prediction has far-reaching economic consequences, from agriculture to transportation, from military deployment to disaster preparedness. Oil exploration is a difficult, often dirty process, but one that is crucial to our fossil-fuel based economies. Doing it better can save enormous investment costs and avoid unnecessary environmental damage. Protein folding is an emerging area and is expected to be important for drug-discovery because it allows predictive modelling, rather than expensive clinical testing, to help determine which drugs will be effective. 

Across all of these different domains, we find that computing power is an important part of the production function. In weather prediction, increasing computing power by 10× improves 3-day-ahead weather predictions by one-third of a degree. Benefits from such improvement would redound broadly.  The literature \cite{teisberg2005economic} estimates that weather prediction is worth \$31.5B\footnote{40.8 B in 2021 dollars.} to the U.S. economy per year \cite{lazo2009300}. In oil exploration, we find substantial differences in the effect that additional computing yields for different companies, perhaps because of differing geologies. But, for example, BP’s drilling success rate goes up by 43 percentage points with each 10× increase in the amount of computing used (but where drilling using the same amount of computing gets harder by 14 ppt per year as easier wells become less available). To put that in context, drilling an onshore well is estimated to cost \$4.9M - \$8.3M depending on the depth \cite{eia2016trends}, so better predictions of which wells will be successful is very valuable. In protein folding, we find that the simulated match to real molecules improves by 6.7 points (in the 0-to-100 GTD-TS score) for each 10x increase in the computing power used. Protein folding is widely expected to dramatically accelerate the pace at which diseases and therapeutics are understood and treated \cite{wang2016exploring} and is estimated to have a market size of \$2.6B in 2021 \cite{imarc2021protein}.

We see similarly large effects from increases in computing power on Chess and Go. In Chess, a 10× increase in computing power correlates with an increase of 242-point Elo -- half the point difference between Masters from Grandmasters. In Go, we find a similar result with a 10x increase in computing power leading to a 246-point Elo improvement.

Perhaps as interesting as our overall findings about the importance of computing, is our ability to contrast the contributions that computing power is making to improved performance to those arising from other sources. Using an analysis-of-variance approach, we find that computing power explains 49-94\% of the variation in output. Put another way, for most of these applications, increases in computing power are at least as important as all other factors put together.

Our findings are particularly important at a time when computer progress is slowing, and thus the ability to get improvements in performance is getting harder \cite{leiserson2020there, thompson2021decline}. Perhaps even more worrisome, if sources of computer improvement (such as Moore’s Law) are running out, then the cost of improvement will rise proportionally to computing power increases. But, since exponential increases in computing power would then come with exponential increases in cost, such improvements are likely to be economically unappetizing, and thus we would expect the rate of progress in many areas to diminish as increases in computing power slow.

\section{Theory}\label{theory}

Production functions, of the type pioneered by Romer \cite{romer_1986}, take the following form:

\begin{equation}
\label{eq1}
    Y = AL^\alpha K^\beta
\end{equation}

Where $Y$ represents output, $K$ represents capital, $L$ labor, and $A$ represents a productivity multiplier that is calculated as a residual. $0<\alpha<1$ and $0<\beta<1$ parameterize the marginal contribution as each input grows. As Syverson \cite{syverson2011determines} has pointed out, this form has the advantage of being a linear approximation to any production function. Two key features of such production functions are that they have decreasing marginal returns in each input, and that they are complementary between inputs. Thus, adding ever more capital becomes less and less efficient, but growing capital and labor proportionally yields mutually reinforcing benefits. 
As I.T. capital becomes increasingly important and evidence mounts that it is complementary, rather than substitutive, to other forms of capital \cite{brynjolfsson_1996}, it makes sense to add it separately as another input factor, to get:

\begin{equation}
\label{eq2}
    Y = AL^\alpha K^\beta IT^\gamma
\end{equation}

For the cases that we will be observing, the rate of change of these inputs will be dramatically different. In particular $\frac{\partial IT}{\partial t}$, $\frac{\partial A}{\partial t}$ \(\gg\) $\frac{\partial L}{\partial t}$,$\frac{\partial K}{\partial t}$, that is, the rate of increase of I.T. capital and of productivity are much higher than those for labor and non-I.T. capital. This is because there has been large exponential growth in the provision of computing power \cite{danowitz2012cpu, leiserson2020there, thompson2021decline} and in the improvement of using that computing power more efficiently through algorithms \cite{holdren2010report, sherry2020fast, fichte2020time} --- this latter term being much faster in specific computing applications than it is for the economy overall. At the same time, there has been relatively little change in labor and non-IT capital\footnote{As an example, the number of employees at the National Oceanographic and Atmospheric Association (NOAA) has decreased from nearly 13,000 in 1997 \cite{united1997budget} to 11,000 in 2015 \cite{wiki2021noaa}}.

These growth rates have two important implications. First, exponential growth in an input can be sufficient to overcome decreasing marginal returns and thus provide a constant (or rising) share of the improvements to output. This can be seen by re-parameterizing IT to show exponential growth over time (i.e. $IT_t=IT_0 \cdot e^{vt}$) as compared to some initial period t=0. Substituting this in yields:  $Y_t=A_tL_t^\alpha K_t^\beta (IT_0 \cdot e^{vt})^\gamma$, where $e^{v}$ is the rate of exponential increase. As this makes clear, if $e^{v \cdot \gamma}>1$ then the decreasing returns to scale are more than overcome by the pace of computing power growth.

The second implication of faster rates of growth for I.T. capital and productivity is that other rate terms are likely to be negligible, i.e.

\begin{equation}
\label{eq4}
	\frac{\partial\mathbf{Y}}{\partial\mathbf{t}} = \frac{\partial\mathbf{Y}}{\partial\mathbf{A}} \cdot
	\frac{\partial\mathbf{A}}{\partial\mathbf{t}}
	+
	\frac{\partial\mathbf{Y}}{\partial\mathbf{IT}} \cdot
	\frac{\partial\mathbf{IT}}{\partial\mathbf{t}}
	+
	\frac{\partial\mathbf{Y}}{\partial\mathbf{L}} \cdot
	\frac{\partial\mathbf{L}}{\partial\mathbf{t}}
	+
	\frac{\partial\mathbf{Y}}{\partial\mathbf{K}} \cdot
	\frac{\partial\mathbf{K}}{\partial\mathbf{t}}
	\approx
	\frac{\partial\mathbf{Y}}{\partial\mathbf{A}} \cdot
	\frac{\partial\mathbf{A}}{\partial\mathbf{t}}
	+
	\frac{\partial\mathbf{Y}}{\partial\mathbf{IT}} \cdot
	\frac{\partial\mathbf{IT}}{\partial\mathbf{t}}
\end{equation}

In the case studies that follow, we estimate how performance (Y) and computing power (IT) have changed over time for these important areas of computing. We find that, as expected, IT has strongly decreasing marginal effects on output, but that $\frac{\partial\mathbf{IT}}{\partial\mathbf{t}}$ has grown so rapidly that $\frac{\partial\mathbf{Y}}{\partial\mathbf{IT}} \cdot \frac{\partial\mathbf{IT}}{\partial\mathbf{t}}$ accounts for most of $\frac{\partial\mathbf{Y}}{\partial\mathbf{t}}$. That is, growth in computing power explains most of the growth in output. In the analysis that follows, we estimate these parameters explicitly, test the robustness of these effects, and consider what these mean for the economic viability of improving outcomes using I.T..

An alternate way of modeling the contribution of I.T. would be via the ideas production function \cite{romer_1989, romer1996advanced, Jones2019}, which models the contribution to knowledge accumulation and productivity improvements, rather than total output. In some ways, this might seem a more natural fit, since we are measuring the skill of systems, which might be imagined as a form of the accumulation of knowledge.  However, ideas production functions implicitly need to model not just performance, but performance per unit of input, and that is not how we measure our outcomes.  Consequently, we use the traditional output-based form of the prodution function for our modeling.

\section{Case Studies}\label{case_studies}
In what follows, we present case studies in five domains about how computing power has been associated with improved performance. In each case, we present the history, summary trends, and correlational evidence. In discussing these, we will use the language of causality, which we will defend in section \ref{sec:analysis} when we analyze these areas collectively.

\subsection{Computer Chess}\label{comp_chess}

Chess is one of the most popular board games in the world and has been played since at least the 6th century A.D. \cite{murray2015history}. The first discussion of chess, from a computational perspective, came in 1950 when the mathematician Claude Shannon published a paper entitled “Programming a Computer for Playing Chess” \cite{shannon1950xxii}. Shortly thereafter, Alan Turing created the first algorithm that a computer could use to play chess. Unfortunately, because of the state of computing at the time, Turing's chess algorithm had to be executed manually. It took 15-30 minutes to calculate each move and the algorithm only considered the consequences of its actions two moves in advance. It could not play a full game, much less beat a professional chess player. According to Kasparov and Friedel, Turing’s algorithm would take less than five milliseconds to calculate each move in a modern computer in 2017 \cite{kasparov2017reconstructing}.  

\begin{figure}[!htbp]
	\centering
	\includegraphics[keepaspectratio=true,scale=0.3]{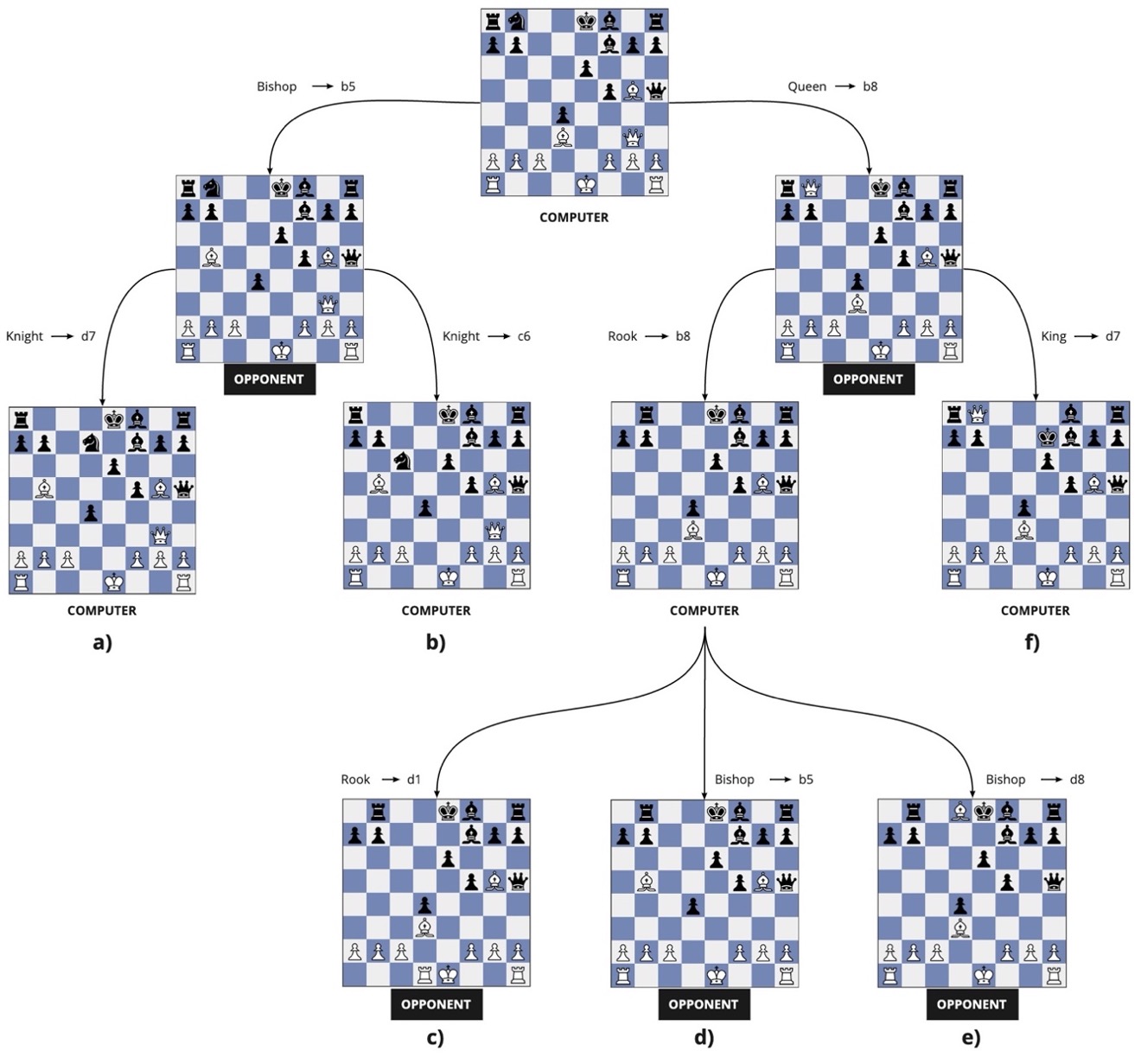}
	\caption{Small subset of a decision tree for chess.}
	\label{chess_board}
\end{figure}

Roughly speaking, all computer chess programs do two types of operations: (1) look ahead to potential future board states and (2) evaluate how likely any board position is to produce a win. For example, a program might look ahead 3 moves (called “plies”) and for each potential result it will evaluate whether it is in a good or bad position by counting when opponent’s pieces can be constrained or taken. In computer science, the resulting analysis is thought of as a tree, where nodes are the board position and edges are the possible moves for each player, as shown in Figure \ref{chess_board} where in d) the computer (white player) checkmates in 3 plies.

Each node has an associated probability of winning. In general, these guide the system towards good moves, but can be misleading if the computer can’t look far enough into the future (and see a potential problem spot) or if it incorrectly evaluates how valuable different board configurations are. Additional computing ameliorates these deficiencies by looking further ahead in the tree or doing better evaluation of a board configuration. Looking farther ahead, however, requires exponentially more computing power since the number of possible moves grows exponentially as you project forward. For example, if each player had 10 potential moves each ply, then there are 10,000 potential evaluations after four plies ($10^{4}$), whereas looking six plies into the future would take 1,000,000 evaluations ($10^{6}$). Smarter evaluation algorithms help this by decreasing the number of moves considered at each point, but at the cost of doing more computation to evaluate each board position. Most programs find a balance between spending time exploring farther ahead and evaluating positions more carefully - although there are exceptions that heavily favor one approach or another\footnote{One such example is AlphaZero, a 2017 program by DeepMind that focuses more on evaluating positions than depth of search.}.

By 1957, Alex Bernstein was able to develop a chess program running on an IBM 704 mainframe that was capable of playing a full game \cite{bernstein1958computer}. Ten years later, the MacHack IV program was the first to play in a tournament. It ended up with a score of one draw and four losses against amateur J.Conroy \cite{wall2008the}. 30 years later, in 1996, after enormous work by the computer chess community Deep Blue, an IBM RS/6000 SP supercomputer capable of calculating 200 million chess positions per second\footnote{By way of comparison, NASA's Pathfinder used the same IBM RS/6000 technology for its onboard flight to Mars \cite{computerworld1997}}, beat the world chess champion, Garry Kasparov. 

Algorithm and hardware improvements have continued to progress since Deep Blue. One recent study showed that a modern computer chess algorithm, running on a single 1994-era 486-DX4 100 MHz machine, would have achieved the Kasparov-levels of performance --- several years ahead of when Deep Blue achieved this with massively more computing power \cite{hippke2020measuring}. On the other hand, old chess programs could also achieve much higher performance running on modern hardware \cite{hippke2021a}. As of July 2021, the best chess program was Stockfish 13. With an Elo score of 3547, Stockfish 14 is 665 points better than the best rating ever achieved by a chess human player, Magnus Carlsen at 2882 in May 2014 \cite{ccrl2021}. This means that the best players in the world are as likely to win a game against Stockfish 13 as a top amateur player would be to win a game against a grandmaster. 

\subsubsection{Computing Power in Computer Chess}

\begin{figure}[!htbp]
	\centering
	\includegraphics[keepaspectratio=true,scale=.75]{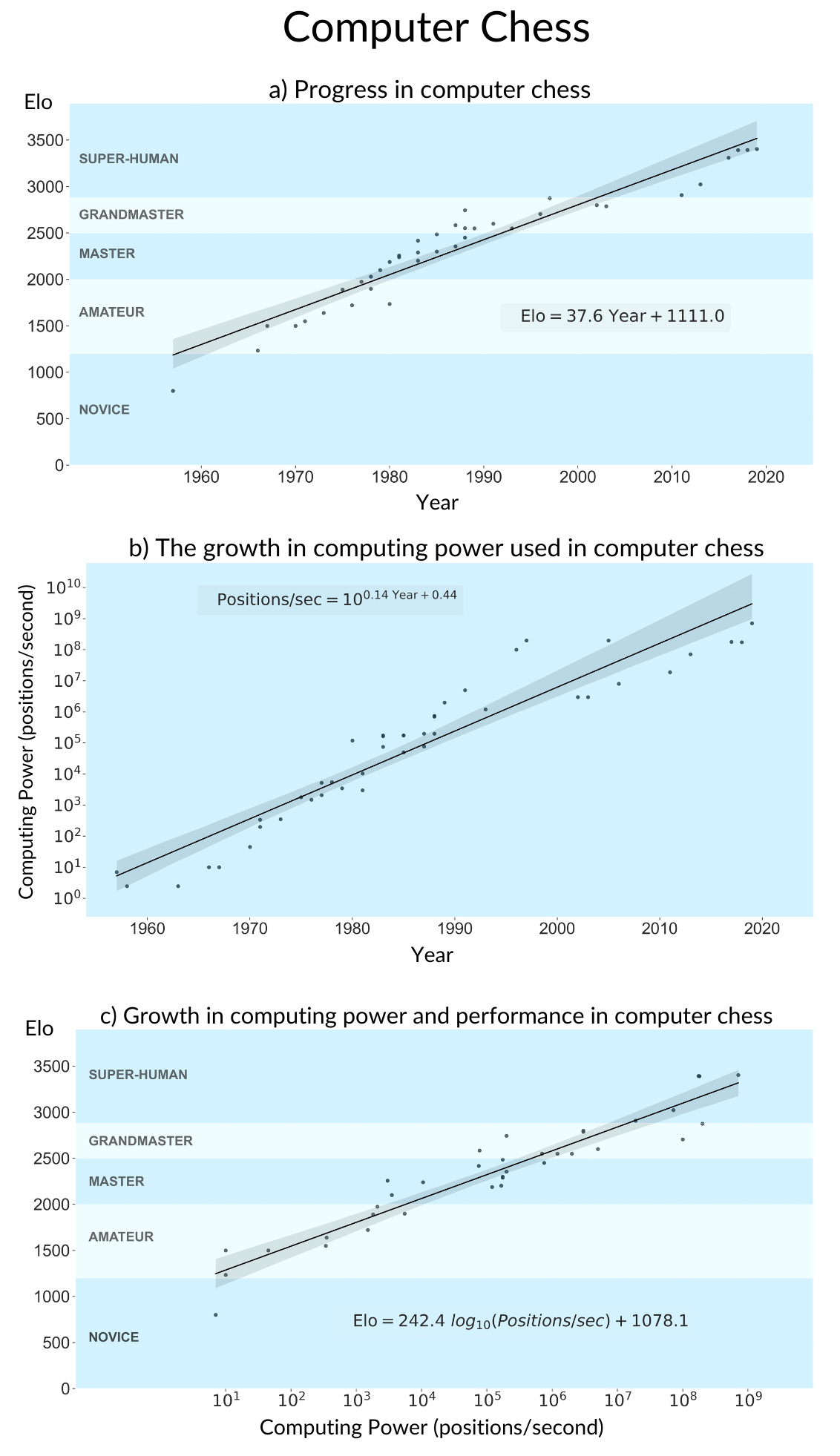}
	\caption{Computer Chess: (a) Elo scores over time, (b) computing power used over time, and (c) Elo scores as computing power increases. In each subfigure, the dots are individual programs, the lines are linear regressions, and the shaded region is the 95\% confidence interval for the regression.}
	\label{chess_img}
\end{figure}

To assess the progress of chess computers over time, we developed an extensive history of computer chess programs from 1957 (Bernstein's program) to 2019 (Komodo 13.1 at the World Computer Chess Championship). For data from 1957 to 2006 we use matches between computers and humans, as gathered from records from international chess associations, research papers, books, databases provided by the community, and others (see Appendix \ref{appendix:data_sources}). From 2006 on we use data from the World Computer Chess Championship, where computers face each other, since at that point computer performance is super-human. Figure \ref{chess_img}a shows how the performance of computer chess has evolved.

Since Bernstein’s 1957 model, computer chess performance went from a novice Elo score to a super-human Elo score of 3547 and increased their computing power usage by a factor of $10^8$. On average, chess programs improved by 37.6 Elo points per year and increased this computing power used by 38\% per year (=$10^{0.14}$).

Analyzing the performance and the number of chess positions shows that an increase of 10× in computing power is associated with an increase in Elo rating of 242 points (statistically significant at the 0.01 level as shown in Table \ref{table:perf_logcp}), as shown in Figure \ref{chess_img}c. This is similar to the estimate made by the Deep Blue team when planning the hardware needed to beat Kasparov “there is a 200-point ELO rating improvement for each order of magnitude improvement in computing speed of the chess machine platform” \cite{tan1995deep}. The variation in computing power explains 88\% of the variation in performance, while the residual variation (e.g. due to algorithmic improvement independent of computing power) only explains 12\%.

\subsection{Computer Go}\label{comp_go}

Go is the oldest board game played in the world, having been invented in China approximately four thousand years ago \cite{bga2020a}. The basic rules of Go are simple: players take turns placing their stones on a 19×19 board and get points by completely surrounding the other player’s stones. This simplicity, however, masks an enormous amount of computational complexity due to the large number of possible moves that players can make each turn. To put this into perspective, after each player makes a single move in chess, there can be 400 possible board configurations, in Go, there can be more than 130,000. Another way of understanding this difference is by comparing the hardware that was used when programs beat the best humans. When Deep Blue beat Garry Kasparov at chess, it used 30 chips containing 480 specialized processors. Twenty years later, when processor chips where ~100× better \cite{hennessy2011computer}, AlphaGo beat Lee Sedol using, roughly, $75\times$ as many processors\footnote{The exact number of processors used by AlphaGo is unknown. Here we make the simplifying assumption that each modern chip has a similar number of processors as the IBM machines (16 processors/chip) and thus AlphaGo’s 1920 CPUs, 280 GPUs, and 48 TPUs \cite{economist2016showdown, silver2017mastering} represent at least $\sim36{,}000$ “processors-equivalents”. In actual fact, the number may be much larger since, for example, a single TPU has $\sim65{,}000$ multiply units, which would make our number an underestimate.}. That is, using $\sim7500\times$+ more computing power.

As with chess, the computational intensity of Go comes from the challenges of looking farther ahead in the game and from evaluating board position. Because of the much larger number of possible moves, the exponential explosion of looking ahead in the game happens much more quickly.

Computer Go began with Albert Zobrist's 1970 dissertation. Programs – such as Interim.2, The Many Faces of Go and Go++ – emerged shortly thereafter. In 2006, an algorithmic improvement\footnote{The introduction of Monte-Carlo tree search (MCTS).} significantly improved performance \cite{gelly2011monte, wall2008the}, allowing a Go computer to achieve a rank of an advanced amateur (1 dan) on a smaller (9×9) board \cite{gelly2012grand}. In 2015, AlphaGo defeated the European Go champion, Fan Hui, in a five-round game by 5-0 \cite{bbc2016google}. This was the first time that an AI system had beaten a human professional player without a handicap. One year later AlphaGo sealed a 4-1 victory over Lee Sedol \cite{theguardian2016alphago}, considered by many to be the best Go player of all time \cite{gibney2016go}.  

\subsubsection{Computing Power in Computer Go}

To assess how Go programs use computing power and how it is impacting their performance we gathered data on computer versus human games. We sourced this data from the British Go Association, the European Go Federation, research papers, tournament and personal reports, books, databases provided by the community, and others (see Appendix \ref{appendix:data_sources}). A key source in this work is the database assemble by Nick Wedd \cite{go2018database}. To fill in missing data, we directly contacted Go developers, teams, professional players and other members of the Go community.

\begin{figure}[!htbp]%
	\centering
	\includegraphics[keepaspectratio=true,scale=.75]{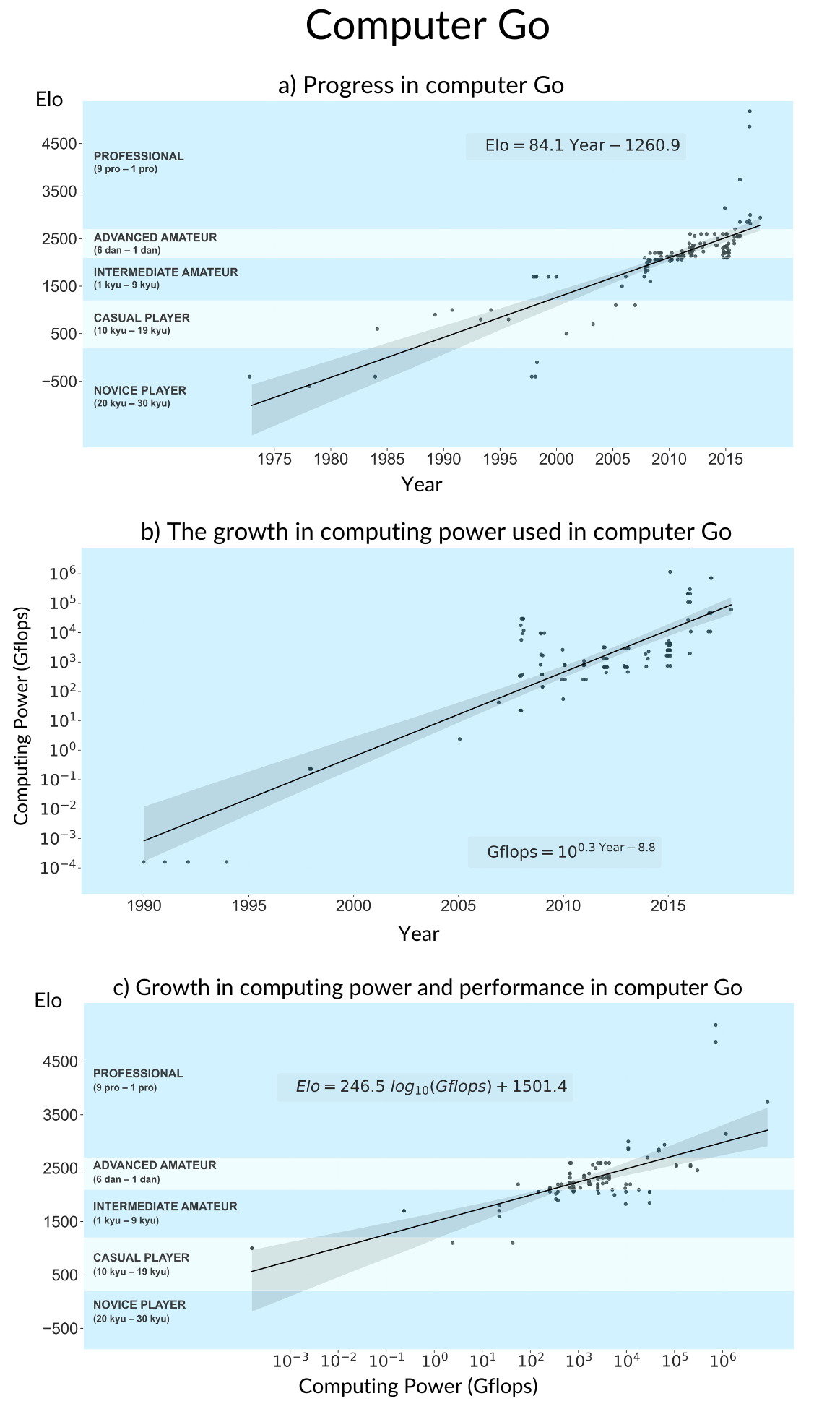}
	\caption{Computer Go: (a) Elo scores over time, (b) computing power used over time, and (c) Elo scores as computing power increases. In each subfigure, the dots are individual programs, the lines are linear regressions and the shaded region represents the 95\% confidence interval for the regression.}
	\label{go_img}
\end{figure}

As Figure \ref{go_img}a shows, there has been enormous improvement in computer Go from the 1970s until today, with an average of 84 Elo points being added to performance per year \footnote{To measure the Elo ratings of Go programs, we use the ranking of the opponent and the number of handicap stones given to infer the rank of the players in kyu/dan, and then use the European Go Federation's conversion from ranks into Elo.}. The best performing systems are now much better than humans, with AlphaGo Zero achieving an Elo of 5135, compared to the human champion Shin Jinseo, who is a 3800 Elo player.

Unfortunately, very little data is available on the computing power used by the early Go systems, so we focus our analysis on the period since 1990, when better data is available. Since that time, the amount of computing power (as measured by floating point operations per second) used by Go programs has increased approximately one hundred billion-fold, a doubling every year (=$10^{0.3}$), as shown in Figure \ref{go_img}b.

Graph (c) compares the growth of computing power in Go with the performance of those programs, revealing a highly significant correlation between them (statistically significant at the p=0.01 level as shown in Table \ref{table:perf_logcp}). In Go, a 10× increase in computing power increases Elo by 246 points on average. In Go, the variation in computing power explains 49\% of the variation in system performance.

\subsection{Weather Forecasting}\label{weather}

It is estimated that 96\% of the U.S. population relies on the weather forecasts provided by the National Oceanic and Atmospheric Association (NOAA), either directly (e.g. through their website) or indirectly via third-parties that process the data that they provide (e.g. the Weather Channel on TV or weather apps like AccuWeather on mobile devices)\cite{lazo2009300, foerster2020what, shepherd2020when}. The economic importance of forecasts is also substantial. Better weather forecasts can guide farmers in when to gather their crops before the first frost, help energy utilities know which generators to have at the ready to heat/cool peoples’ homes, and can advise people to take shelter if a hurricane is headed their way. These, and the many other ways that forecasts guide people in the economy, are estimated to be worth approximately \$31.5 billion per year \cite{lazo2009300}.

Before the mid-nineteenth century, weather forecasting was done based on weather lore, personal observations and simple measurements (e.g. of humidity) \cite{graham2002weather}. Once the telegraph was invented, observers in different locations communicated to produce the earliest weather maps. Today, there is a vast web of sensors that perform these tasks, including over 10,000 manned and automatic surface weather stations, 1,000 upper-air stations (e.g. weather balloons), 7,000 ships, 100 moored and 100 drifting buoys, hundreds of weather radars and 3,000 specially equipped commercial aircraft, together with 30 meteorological and 200 research satellites \cite{wmo2020observations}. These observations are sent to supercomputers that estimate enormous systems of equations to produce predictions. Flynn \cite{flynn2018forecasts} estimates that even the earliest numerical weather forecast models would require 204,800 people to solve manually.

The computational burden of numerical weather prediction can largely be explained by three factors: resolution, dynamics and variations. The resolution of the model refers to the geographic space modeled as a single unit. For example, the current U.S. model divides the atmosphere into chunks that are 13km long $\times$ 13km wide \cite{NWS2020doc}. The resolution of models has changed dramatically over time. For example, the global leader in weather forecasting models is the European Centre for Medium-Range Weather Forecasts (ECMWF) \cite{states2013restoring}, which has increased in resolution from 210 kilometers in 1979 down to a 9 kilometers in 2016 \cite{ecmwf2015forty, schulthess2018reflecting}. Improvements in these three dimensions of resolution are computationally expensive, with each doubling requiring an ($2^3$) 8-fold rise in the amount of computation \cite{neumann2019assessing}. A common goal in the weather prediction community is to get models to 1 km resolution with 200 vertical levels  and 100 variables for the dynamics, which would help accurately model mountain ranges and other important features but would require 2,000 times as much computing power \cite{bauer2016today}. 

The dynamics of weather prediction refer to how different units of space (of whatever resolution) interact with each other and how this is modeled. Many different techniques are used for this, for example using partial differential equations of physical interactions to model variables like temperature, wind, pressure, and other hydrodynamics and thermodynamics variables \cite{lynch2006emergence}. The models take the initial state of the model and predict what will happen in 24, 48, 72, or more hours. Because small variations in the initial conditions can produce large differences in outcomes, weather prediction often uses ensemble methods, running the model many times with slightly perturbed initial conditions in order to explore the variation in outcomes that this produces. This can easily increase the amount of computing needed for a calculation by a factor of 50 to 100 \cite{bauer2016today}.

\subsubsection{Computing Power in Weather Forecasting}

To assess progress in temperature prediction, we analyze data provided by NOAA on the performance of their high-performance computing systems since the 1950s. We measure performance with the mean absolute error (in Fahrenheit degrees) between the temperature prediction and the actual temperature \footnote{Consistent with the norms in this field, only the error in the prediction of maximum and minimum temperature is shown, but this result holds when we use other temperature indicators such as average temperature. Trends in hurricaine prediction, another area that NOAA predicts, show similar progress}. As shown in Figure \ref{weather}(a), the errors made by NOAA in predicting temperatures have fallen dramatically since the 1970s. For example, predictions for weather 3 days in the future have dropped from an error of 5.8 degrees Fahrenheit in 1972 to 3.0 degrees in 2017 a drop of 47\%.

\begin{figure}[!htp]
	\centering
	\includegraphics[keepaspectratio=true,scale=.7]{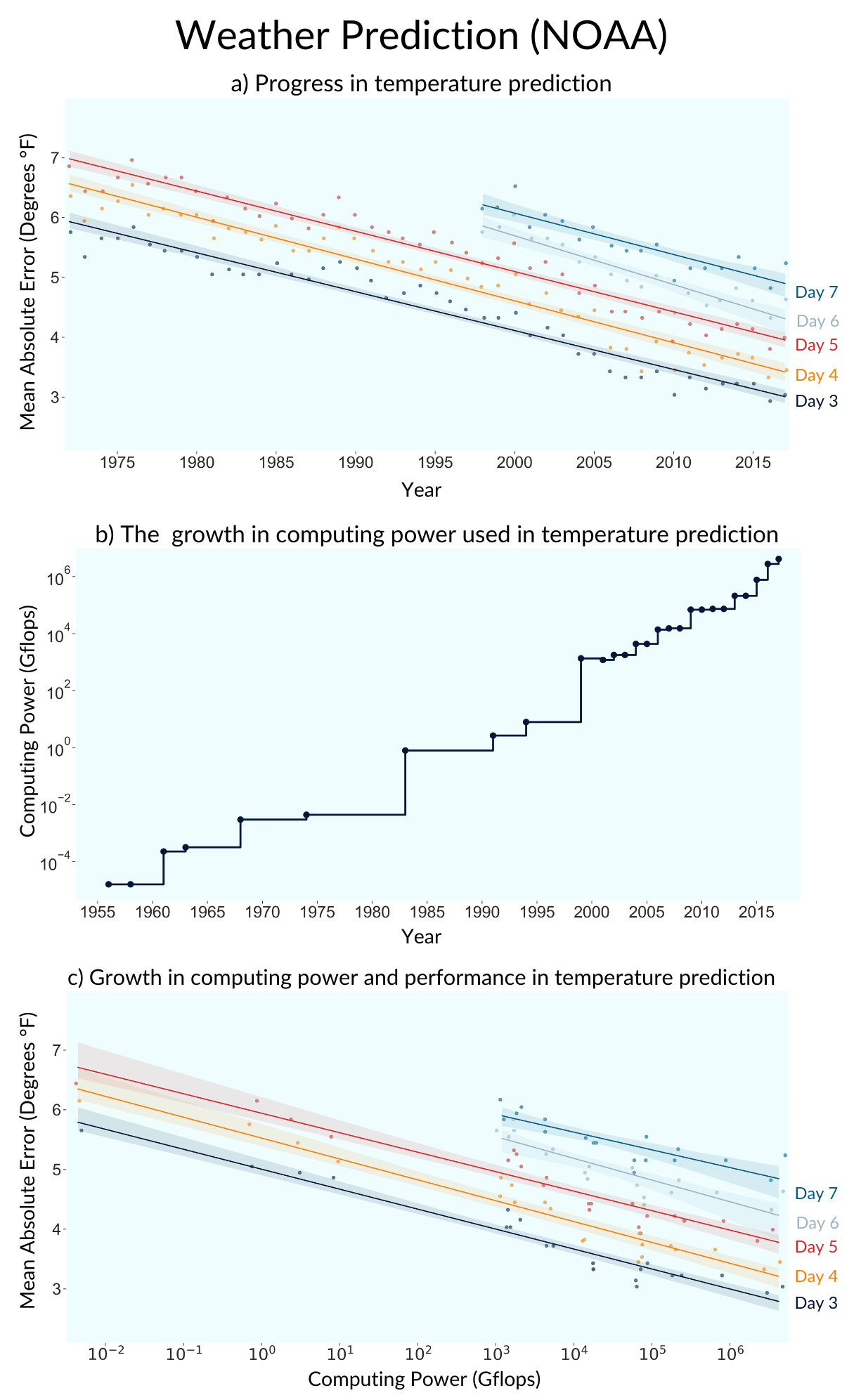}
	\caption{Temperature Prediction: (a) Forecast accuracy over time, (b) Computing power over time, as measured by system capacity, and (c) forecast accuracy as computing power increases. Lines with shaded regions are linear regressions with 95\% confidence intervals \cite{WPCV2020}}.
	\label{weather}
\end{figure}

%is understood to be a good indication of the overall forecast ability \cite{}.} 
The computing power used by NOAA for weather prediction has also escalated dramatically, growing nearly a trillion-fold from 1956 until 2017, as shown in Figure \ref{weather}(b), representing an increase of 48.2\% per year.\footnote{Note: unlike for Chess and Go, this measurement is about the functionality of the whole system, not the computation needed for a particular task. That said, discussions with practitioners indicate that the time window used for calculations have been quite stable due to reporting requirements (e.g. updating forecasts hourly) and thus the two should be highly correlated.}  The large jumps in this figure represent new supercomputers coming online, whereas the smaller jumps come from expansions to existing systems.

Combining the data from Figure \ref{weather}(a) and Figure \ref{weather}(b) allows us to examine the progress in prediction as computing power has varied. Figure \ref{weather}(c) shows the resultant tight correlation. In weather prediction, increasing computing power by 10× yields a decrease in prediction error of 1/3 of a degree Fahrenheit (statistically significant at a p-value $<$ 0.01 (see Table \ref{table:perf_logcp}). Perhaps even more notable, variation in the amount of computing power used explains 73-94\% of the variation in performance, suggesting that improvements in computing power (and in the algorithms needed to harness it) are responsible for the vast majority improvements in weather prediction performance.

\subsection{Protein Folding}\label{protein}

Proteins are the molecular machines of the body, performing virtually all the important biochemical processes \cite{zhang2008progress}. When first produced in the body, proteins consist of a linear chain of units\footnote{Amino acid residues.} which then fold into a three-dimensional structure that interacts with the world. Biologists care a great deal about protein folding because the folding pattern determines a protein’s structure, what it binds to, its stability, and other properties that are important for health and medicine \cite{wright1999intrinsically}. An early discussion of predicting protein folding highlighted the incredible difficulty of these calculations: a moderate-sized protein might have $10^{300}$ possible configurations \cite{levinthal1969mossbauer}. 

Assessments of protein folding skill happen at events such as the Critical Assessment of Protein Structure Prediction Experiments (CASP), where hundreds of protein sequences are presented to the computational modelling community who attempt to predict their folded structure. Success is measured with a GDT\_TS (Global Distance Test\_Total Score) which calculates the similarity between the results of protein structure prediction and the experimentally determined structure spotted by NMR or X-RAY crystallography. The score ranges from 0 when the two structures are completely different to 100 when the predicted structure is exactly the actual structure. In the analysis that follows we focus on GDT\_TS scores for free modelings problems, where the full folding problem must be solved (there are also simpler contests where similar sequences can be used as templates, vastly simplifying the difficulty of the problem).  

\subsubsection{Computing Power in Protein Folding}

Tracking the effects of computing power is significantly harder in protein folding than other domains. Because, while it is easy to source information from CASP on how good structural predictions are \cite{CASPweb}, virtually none of the related papers on these models report their computing power usage\footnote{Either directly or via a measure of computing hardware burden \cite{thompson2020computational}, calculated as \# processors × Computation Rate × time.}. After carefully reviewing all the publications from CASP 7 to CASP 13, around 200 papers in total, we are only able to extract computation data for 5, including AlphaFold1\footnote{The computing power for Alphafold1 is estimated as 4 times the computation used in ProSPr (Billings, 2019). This estimation might miss the massive training size and potential ensembles embedded in the original work of Alphafold1.} and AlphaFold2. Fortunately, these models span a 100,000 fold difference in computing power so we are still able to see significant variation, as shown in Figure \ref{protein}. 

\begin{figure}[!htp]
	\centering
	\includegraphics[keepaspectratio=true,scale=.75]{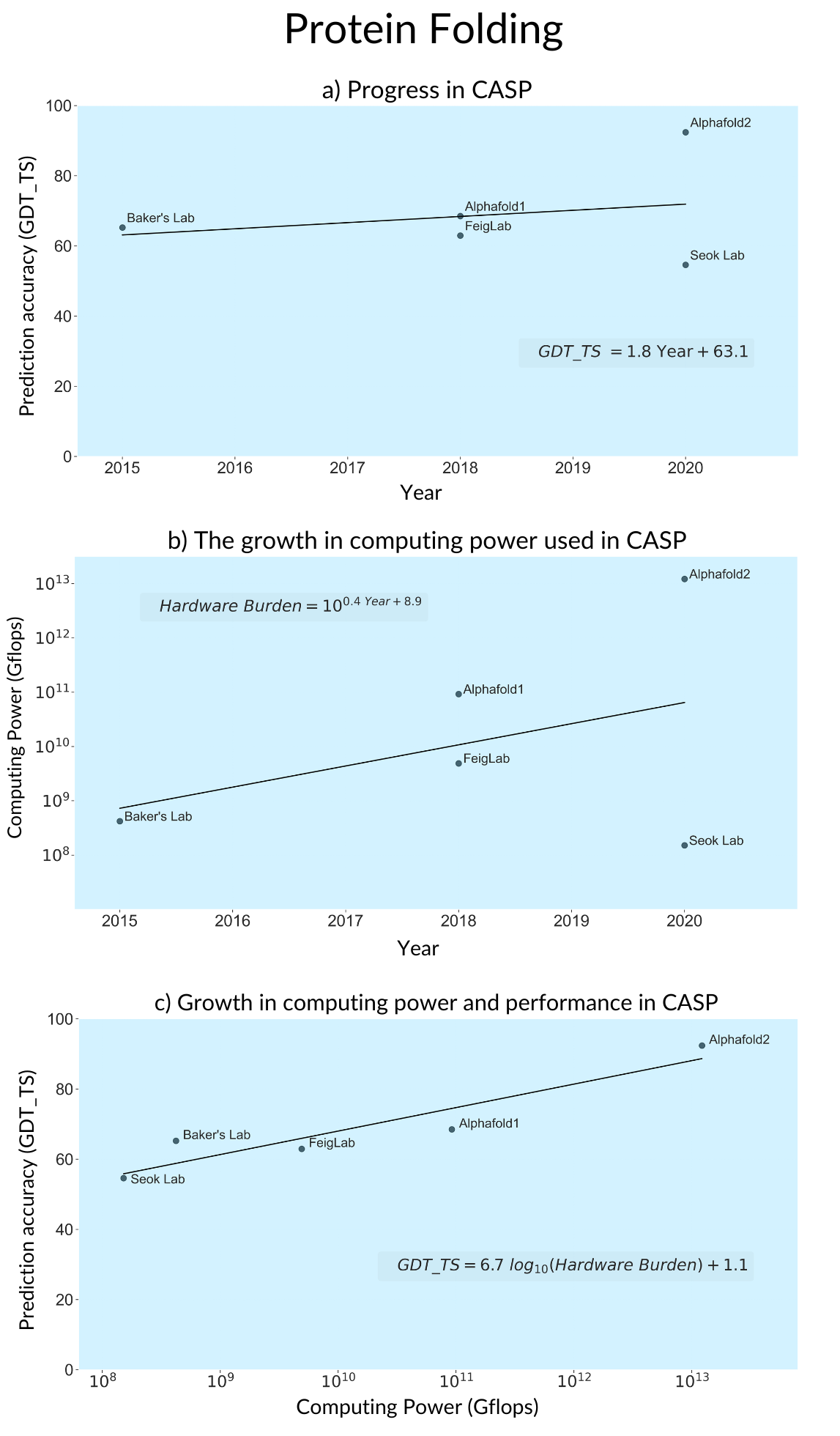}
	\caption{Protein Folding: (a) Prediction accuracy (as measured by GDT\_TS scores) over time, (b) computing power used over time, and (c) Prediction accuracy (GDT\_TS scores) as computing power increases. All data from CASP competitions. Lines are linear regressions, shaded areas are their 95\% confidence intervals.}
	\label{protein}
\end{figure}

Overall, we observe rapid progress. In the six years from 2015 to 2021 performance has risen from 65 to 69, or 92 . This has been accompanied by a huge increase in computing power being used, from 421M Gflops to 12.2 trillion Gflops, a compound rate of increase of $4\times$ per year. We also find that computing power changes explain 87\% of the variation, with each 10x increase in computing power being associated with an 6.7 increase GDT\_TS performance.

Since this analysis is based on only a small number of data points, we also conduct an alternative empirical analysis in Appendix \ref{appendix:protein}. That also finds strong support for role of computing power in performance increases.

\subsection{Oil Exploration}\label{oil}

Drilling for oil is expensive, costing \$3.3 trillion worldwide yearly \cite{investopedia2020what} and representing ~8.5\% of the capital expenditures for a major oil company such as Chevron (Chevron, 2019). Drilling an on-shore well is estimated to cost between \$4.9 M to \$8.3 M on average \cite{eia2016trends}, and off-shore wells are estimated to cost \$650 M on average \cite{harvey2020what}. With such high costs, there is great value in assuring that drilling produces active wells and not ‘dry holes’. Computational methods can be of great value in guiding drilling. The probability of drilling a successful well is 70\% with 3D seismic methods, while 30\%-35\% without 3D \cite{rundle13}.

Seismic modelling emerged in the 1950s. It models the physical structure of the earth by understanding how seismic waves (e.g. concussions done by the mappers themselves) travel and reflect off of different materials. Computationally, this involves solving a series of wave equations that contain important physical parameters of the geologic materials. Unfortunately, even in the late 1970s, computing power was insufficient for solving the comprehensive set of wave equations in more complex seismic models \cite{bamberger1982inversion}. Instead, data was gathered in two dimensions, with repeated sampling needed to form a set of slices spanning a volume of the subsurface.

Understanding improved when modeling was done in 3D, yielding better resolution and structural information \cite{davies20043d}. However, adding dimensions also means adding parameters, which is computationally expensive \cite{karavaev2015technology}. As the earliest literature that described this method said, “because of the large number of unknowns…the calculation…by finite-differences using small increments of each of the variables…would lead to many hundreds of evaluations…hence to many hundreds of solutions of the wave equation…This would be computationally prohibitive.” \cite{bamberger1982inversion}, Fortunately, as the cost of computing power fell and available computing power grew, 3D seismic modeling became possible by 1982. By 1991, this was further extended to include time-lapse (so-called “4D modeling”).

\subsubsection{Computing Power in Oil Exploration}

According to the Energy Information Administration, the rate of drilling success (i.e. not getting dry wells) in the United States improved from ~10\% in the 1940s to 70\% in the 2010s \cite{EIA_energy2012}. To understand how drilling performance has improved, we considered the computing power used by the largest oil companies around the world: Total, ExxonMobil, Chevron, and BP. Reflecting the high value of seismic modeling, each of these companies has a supercomputer more powerful than the supercomputer that NOAA uses for weather prediction for the whole country. And, in fact, another petroleum company, Eni, has the world’s most powerful supercomputer in commercial use - it has 3,200 Nvidia Tesla GPUs and 18.6 petaflops operational capacity \cite{hpc2018eni}. Companies also greatly increased the power of their computers over time: Exxon Mobil from $\sim79$ Gflops in 1998 to 26,000,000 Gflops in 2019 (330K-fold increase), Chevron from 75 Gflops in 1991 to 2,000,000 Gflops in 2012 (a 27K-fold increase), and BP from $\sim100$ Gflops in 1999 to 20,000,000 Gflops in 2020 (a 200K-fold increase).

Using data from iHS Markit energy portal U.S. Data Online, a comprehensive well archive with reports on over 4 million wells and 2.7 million producing entities, we analyzed the success that BP had in exploratory drilling for offshore wells \cite{ihs2020what}. We focus on off-shore wells because these are the ones where seismic is the main investigation method, and of those we focus on wildcat wells which are exploratory (rather than drilling into a known area). Figure \ref{oil} presents the drilling success rates for these wells over time, after removing the time trend because oil drilling gets progressively more difficult as easier areas are exhausted.

\begin{figure}[!htbp]
	\centering
	\includegraphics[keepaspectratio=true,scale=.9]{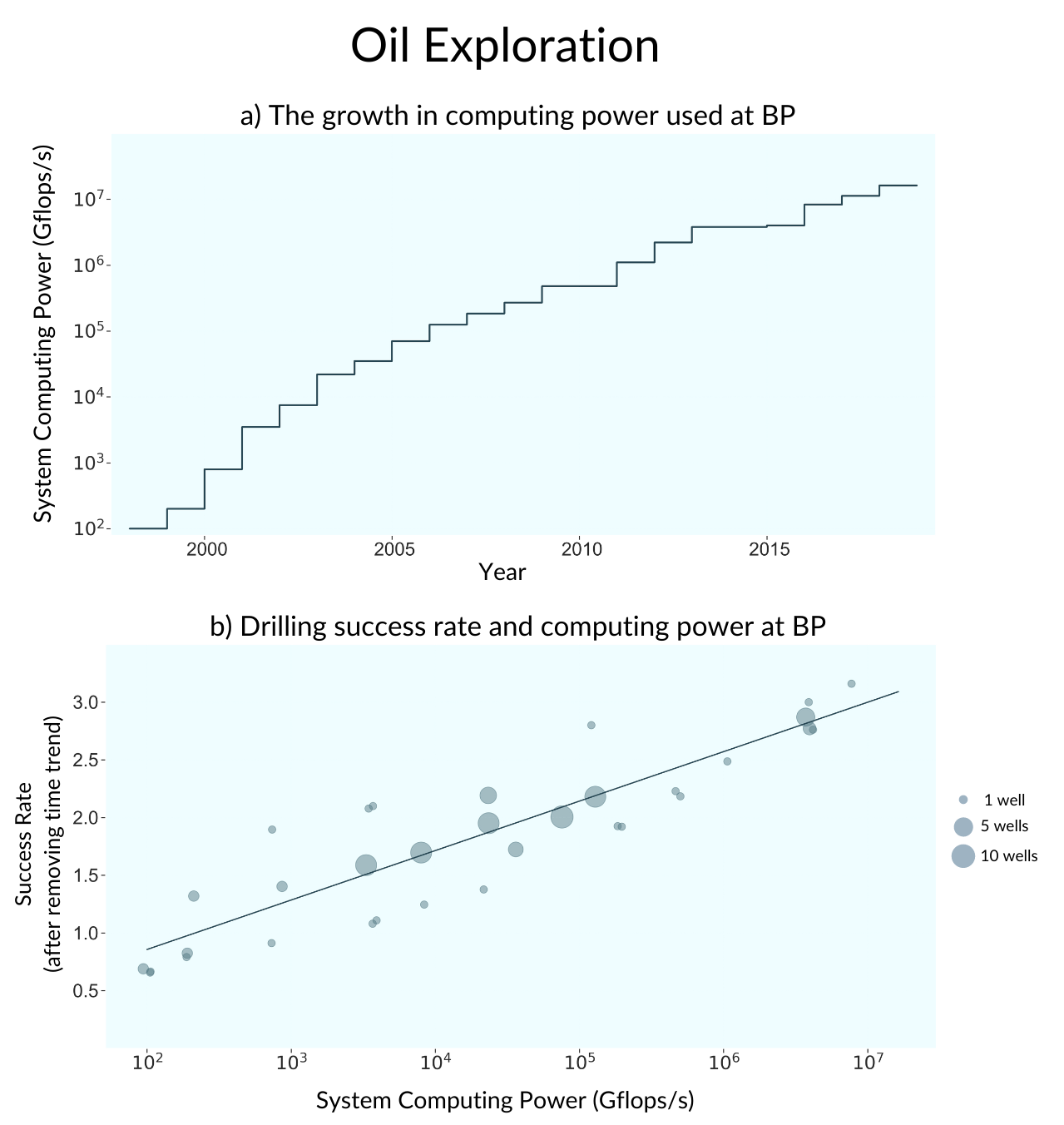}
	\caption{Oil Exploration: a) The growth in computing power used for oil exploration at BP. b) Increase in the drilling success rate at BP as computing power used increases. The success rate shown is after excluding the overall time trend (which is why it can exceed 1). Each circle represents the number of wells in a specific year at a specific depth interval.}
	\label{oil}
\end{figure}

These data show that a 10× increase in computing power is been associated with over 43 percentage point improvement in the drilling success rate for BP (significant at a p-value of 1\%). Importantly, this improvement happens against a backdrop of drilling getting harder as easier locations have already been drilled --- that is, over time the drilling success rate with a given amount of computing power is falling.  After de-trending these rates, computing power explains 85\% of the variation in the drilling success rate, affirming the importance of computing power in this area. 

There are important caveats to this finding. First, our intention had been to analyze four companies in our analysis: BP, Chevron, ExxonMobil and Total. We excluded ExxonMobil and Total for having too few data points, and Chevron because we did not trust the data quality after observing a giant, unexplained drop in its drilling success in 1992, as well as frequent year-to-year jumps between a 0 and 100\% success rate which seemed implausible. Second, because we are looking at the de-trended data, we are only explaining the residual variance after that detrending, not all the variance.

\subsection{Summary of dependence on computing power}

Looking across each of these five domains, we find (Table \ref{table:perf_logcp}) the following relationships between computing power and additional performance:

\begin{table}[!htp] \centering 
\caption{Performance over computation in logarithm scale.}
\label{table:perf_logcp} 
\resizebox{\textwidth}{!}{
 
\begin{tabular}{@{\extracolsep{5pt}}lcccccc} 
\\[-1.8ex]\hline 
\hline \\
 & & 
   \multicolumn{1}{c}{\begin{tabular}[c]{@{}c@{}}Computer Chess\end{tabular}} &
   
   \multicolumn{1}{c}{\begin{tabular}[c]{@{}c@{}}Computer Go\end{tabular}} &

   \multicolumn{1}{c}{\begin{tabular}[c]{@{}c@{}}Oil Exploration\\ (BP)\end{tabular}} &
   
   \multicolumn{1}{c}{\begin{tabular}[c]{@{}c@{}}Protein Folding\\ (CASP)\end{tabular}} \\ \\
   
\cline{3-6} \\

 & & 
 \multicolumn{1}{c}{\begin{tabular}[c]{@{}c@{}}ELO\end{tabular}}  &
 
 \multicolumn{1}{c}{\begin{tabular}[c]{@{}c@{}}ELO\end{tabular}}  &
 
\multicolumn{1}{c}{\begin{tabular}[c]{@{}c@{}}Drilling\\ Success\ $\%$\end{tabular}} &

 \multicolumn{1}{c}{GDT\_TS}\\ & & 
 \multicolumn{1}{c}{(1)} &
 \multicolumn{1}{c}{(2)} &
 \multicolumn{1}{c}{(3)} &
 \multicolumn{1}{c}{(4)} \\ \\
 
\hline \\

 $Constant$ & & 
 \multicolumn{1}{c}{1,078$^{***}$} &
 \multicolumn{1}{c}{\ 1,501$^{***}$} &
 \multicolumn{1}{c}{ \ \ \ \ 0.00002} &
 \multicolumn{1}{c}{\ 1.11} \\
 
  & & 
  \multicolumn{1}{c}{(82)} &
  \multicolumn{1}{c}{(87)} &
  \multicolumn{1}{c}{(0.15)} &
  \multicolumn{1}{c}{(15.05)} \\ 
  
 $log_{10}(Computing\ Power)$ & & 
 \multicolumn{1}{c}{\ \ \ 242$^{***}$} &
 \multicolumn{1}{c}{\ \ \ 246$^{***}$} &
 \multicolumn{1}{c}{\ \ \ 0.43$^{***}$} &
 \multicolumn{1}{c}{\ \ \ \ 6.69$^{**}$} \\ 
 
  & & 
  \multicolumn{1}{c}{(16)} &
  \multicolumn{1}{c}{(24)} &
  \multicolumn{1}{c}{(0.03)} &
  \multicolumn{1}{c}{ \ (1.47)} \\ \\
  
\hline 
\\ [-1.8ex]
Observations & & 
\multicolumn{1}{c}{31} &
\multicolumn{1}{c}{104} &
\multicolumn{1}{c}{33} &
\multicolumn{1}{c}{5} \\ \\

R$^{2}$ & & 
\multicolumn{1}{c}{0.88} &
\multicolumn{1}{c}{0.49} &
\multicolumn{1}{c}{0.85} &
\multicolumn{1}{c}{0.87} \\ \\

Adjusted R$^{2}$ & &
\multicolumn{1}{c}{0.88} &
\multicolumn{1}{c}{0.49} &
\multicolumn{1}{c}{0.85} &
\multicolumn{1}{c}{0.83} \\ \\

Residual Std. Error & & 

   \multicolumn{1}{c}{\begin{tabular}[c]{@{}c@{}}180.33 \\ df=29\end{tabular}} &
   \multicolumn{1}{c}{\begin{tabular}[c]{@{}c@{}}395.54 \\ df=102\end{tabular}} &
   \multicolumn{1}{c}{\begin{tabular}[c]{@{}c@{}}0.37 \\ df=31\end{tabular}} &
   \multicolumn{1}{c}{\begin{tabular}[c]{@{}c@{}}5.81 \\ df=3\end{tabular}} \\ \\

F Statistic 
& & 

   \multicolumn{1}{c}{\begin{tabular}[c]{@{}c@{}}218$^{***}$ \\ df=1; 29\end{tabular}} &
   \multicolumn{1}{c}{\begin{tabular}[c]{@{}c@{}}218$^{***}$ \\ df=1; 102\end{tabular}} &
   \multicolumn{1}{c}{\begin{tabular}[c]{@{}c@{}}182$^{***}$ \\ df=1; 31\end{tabular}} &
   \multicolumn{1}{c}{\begin{tabular}[c]{@{}c@{}}20$^{**}$ \\ df=1; 3\end{tabular}} \\ 
   
   \hline \\
 & & & \multicolumn{1}{c}{Weather Prediction} & &  \\ \\

\cline{2-7} \\

 & 
 \multicolumn{1}{c}{Day\ 3} &
 \multicolumn{1}{c}{Day\ 4} &
 \multicolumn{1}{c}{Day\ 5} &
 \multicolumn{1}{c}{Day\ 6} &
 \multicolumn{1}{c}{Day\ 7} \\ 
 
 & 
 \multicolumn{1}{c}{(4)} &
 \multicolumn{1}{c}{(5)} &
 \multicolumn{1}{c}{(6)} &
 \multicolumn{1}{c}{(7)} &
 \multicolumn{1}{c}{(8)} \\ \\
 
\hline \\

 $Constant$ & 
 \multicolumn{1}{c}{5.00$^{***}$} &
 \multicolumn{1}{c}{5.53$^{***}$} &
 \multicolumn{1}{c}{5.94$^{***}$} &
 \multicolumn{1}{c}{6.64$^{***}$} &
 \multicolumn{1}{c}{6.81$^{***}$} \\
 
  & 
  \multicolumn{1}{c}{(0.08)} &
  \multicolumn{1}{c}{(0.11)} &
  \multicolumn{1}{c}{(0.09)} &
  \multicolumn{1}{c}{(0.24)} &
  \multicolumn{1}{c}{(0.21)} \\
  
 $log_{10}{(Computing\ Power)}$ &
 \multicolumn{1}{c}{-0.33$^{***}$} &
 \multicolumn{1}{c}{-0.35$^{***}$} &
 \multicolumn{1}{c}{-0.33$^{***}$} &
 \multicolumn{1}{c}{-0.36$^{***}$} &
 \multicolumn{1}{c}{-0.30$^{***}$} \\
 
  & \multicolumn{1}{c}{(0.02)}
  & \multicolumn{1}{c}{(0.02)}
  & \multicolumn{1}{c}{(0.02)}
  & \multicolumn{1}{c}{(0.05)}
  & \multicolumn{1}{c}{(0.04)} \\ \\
  
\hline 
\\ [-1.8ex]
 Observations & 
 \multicolumn{1}{c}{22} &
 \multicolumn{1}{c}{22} &
 \multicolumn{1}{c}{22} &
 \multicolumn{1}{c}{18} &
 \multicolumn{1}{c}{18} \\ \\
 
 $R^2$ &
 \multicolumn{1}{c}{0.94} &
 \multicolumn{1}{c}{0.90} &
 \multicolumn{1}{c}{0.92} &
 \multicolumn{1}{c}{0.76} &
 \multicolumn{1}{c}{0.73} \\ \\
 
 Adjusted $R^2$ &
 \multicolumn{1}{c}{0.94} &
 \multicolumn{1}{c}{0.90} &
 \multicolumn{1}{c}{0.91} &
 \multicolumn{1}{c}{0.74} &
 \multicolumn{1}{c}{0.71} \\ \\
 
 Residual Std. Error & 
 
 \multicolumn{1}{c}{ \begin{tabular}[c]{@{}c@{}}0.19 \\ df=20\end{tabular}} &
 
  \multicolumn{1}{c}{ \begin{tabular}[c]{@{}c@{}}0.26 \\ df=20\end{tabular}} &
 
  \multicolumn{1}{c}{\begin{tabular}[c]{@{}c@{}}0.22 \\ df=20\end{tabular}} &
 
  \multicolumn{1}{c}{\begin{tabular}[c]{@{}c@{}}0.23 \\ df=16\end{tabular}} &
  
   \multicolumn{1}{c}{\begin{tabular}[c]{@{}c@{}}0.21 \\ df=16\end{tabular}} & \\ \\
   
 F Statistic 
 & 
 
   \multicolumn{1}{c}{\begin{tabular}[c]{@{}c@{}} 309$^{***}$ \\ df=1; 20\end{tabular}} &
 
    \multicolumn{1}{c}{\begin{tabular}[c]{@{}c@{}} 188$^{***}$ \\ df=1; 20\end{tabular}} &

   \multicolumn{1}{c}{\begin{tabular}[c]{@{}c@{}} 220$^{***}$ \\ df=1; 20\end{tabular}} &
   
   \multicolumn{1}{c}{\begin{tabular}[c]{@{}c@{}} 50$^{***}$ \\ df=1; 16\end{tabular}} &

   \multicolumn{1}{c}{ \begin{tabular}[c]{@{}c@{}} 43$^{***}$ \\ df=1; 16\end{tabular}}  \\ 
\hline
\hline \\ \\ [-1.8ex]
\textit{Note:} & \multicolumn{5}{r}{$^{*}$p$<$0.1; $^{**}$p$<$0.05; $^{***}$p$<$0.01}
\end{tabular}

}
\end{table}

\section{Analysis and Discussion}\label{sec:analysis}

Having presented our case studies graphically, we now explicitly estimate the four key parameters from our theory model, in particular: $\psi=\frac{\partial IT}{\partial t}$ the rate of increase of computing power, $\gamma$ the exponent for I.T., $\frac{\partial Y}{\partial IT} \cdot \frac{\partial IT}{\partial t}$ the proportion of improvement in performance explainable from improvement in I.T., and $\rho \equiv \frac{\partial Y}{\partial t} - \frac{\partial Y}{\partial IT} \cdot \frac{\partial IT}{\partial t}$ the residual portion of growth not explained by increases in computing power.  

Throughout this section, we shall use the language of causality in interpreting our results. Normally with time series data, this would be problematic without a natural experiment, instrumental variable or other means of providing statistical identification. But here, we rely not on statistical analysis of our data to get causality, but on the extensive scientific experimentation done by those in these fields. That is, in all five of these areas, engineering and scientific understanding has been built using experiments that prove that computing power causally improves performance (e.g. \cite{director2010increasing, neumann2019assessing}). For example, NOAA does tests showing that weather forecasts can be improved by running a for a longer time than it would otherwise for a calculation. For example, to test whether a $2\times$ increase in computing would give a better result, they could run a calculation that needs to be done in 1 hour for 2 hours. The results from this testing are then part of the funding approval process for getting sufficient computing power to run the new method in the required amount of time. Thus, it is the experimental testing in these fields, not after-the-fact statistical identification, that we use to get causality\footnote{One potential caveat to this is chess in the period since 1997. Since that time, the cost of computing being used has fallen, suggesting that there may have been less vigilance on the budget and thus on proving additional performance.}. 

The economic cost of the computing systems used provides a second argument for why the causality runs from computing power to performance: revealed preference. Even after accounting for rapid hardware improvement rates, all of these areas have shown enormous increases in the cost of the computing power being used. Literally, hundreds of millions of dollars have been spent on these systems because their owners were convinced that causality runs in this direction. Had this not been true, we would have expected the owners of these systems to update their computers to get the benefit of newer hardware while maintaining or lowering costs. 

While we find the experiments conducted by scientists to be particular compelling evidence of causality (much more so that typical statistical measures), we also recognize that this way of establishing it is unorthodox in this literature. As such, in Appendix \ref{appendix:first_diff}, we also present more-traditional statistical tests for establishing causality.

\subsection{The Growth in Computing Power}

% stopped here ---------------------------------------------------------------------------

Table \ref{logcp_year} shows our estimates for $\psi$ the rate of increase of computing power across our domains:

\begin{table}[!htp] \centering 
\caption{Growth in computing power per year.}
\label{logcp_year}
\resizebox{\textwidth}{!}{

\begin{tabular}{@{\extracolsep{5pt}}lD{.}{.}{-4} D{.}{.}{-4} D{.}{.}{-4} D{.}{.}{-4} D{.}{.}{-4}} 
\\[-1.8ex]\hline 
\hline \\

 & 
 \multicolumn{1}{c}{\begin{tabular}[c]{@{}c@{}} Computer Chess\end{tabular}} &
 \multicolumn{1}{c}{\begin{tabular}[c]{@{}c@{}} Computer Go\end{tabular}} &
 \multicolumn{1}{c}{\begin{tabular}[c]{@{}c@{}} Weather Prediction\end{tabular}} &
 \multicolumn{1}{c}{\begin{tabular}[c]{@{}c@{}} Oil Exploration\\ (BP)\end{tabular}} &
 \multicolumn{1}{c}{\begin{tabular}[c]{@{}c@{}} Protein Folding\\ (CASP)\end{tabular}} \\ \\
 
 \cline{2-6} \\
 
 &
 \multicolumn{1}{c}{\begin{tabular}[c]{@{}c@{}} Positions/sec\\ ($log_{10}$)\end{tabular}} &
 \multicolumn{1}{c}{\begin{tabular}[c]{@{}c@{}} Gigaflops\\ ($log_{10}$)\end{tabular}} &
 \multicolumn{1}{c}{\begin{tabular}[c]{@{}c@{}} Gigaflops\\ ($log_{10}$)\end{tabular}} & 
 \multicolumn{1}{c}{\begin{tabular}[c]{@{}c@{}} Gigaflops\\ ($log_{10}$)\end{tabular}} &
 \multicolumn{1}{c}{\begin{tabular}[c]{@{}c@{}} Hardware Burden\\ ($log_{10}$)\end{tabular}} \\ &
 \multicolumn{1}{c}{(1)} & \multicolumn{1}{c}{(2)} & \multicolumn{1}{c}{(3)} & \multicolumn{1}{c}{(4)} & \multicolumn{1}{c}{(5)}\\ \\
 
\hline\\

 $Constant$ & 
 \multicolumn{1}{c}{0.44$^{***}$}  &
 \multicolumn{1}{c}{-8.81$^{***}$} &
 \multicolumn{1}{c}{-5.22$^{***}$} &
 \multicolumn{1}{c}{2.78$^{***}$} & 
 \multicolumn{1}{c}{8.86}\\ 
 
  & \multicolumn{1}{c}{(0.27)} 
  & \multicolumn{1}{c}{(0.69)}
  & \multicolumn{1}{c}{(0.17)}
  & \multicolumn{1}{c}{(0.15)}
  & \multicolumn{1}{c}{(1.88)}\\ 
  
 $Year$ &
 \multicolumn{1}{c}{0.14$^{***}$} &
 \multicolumn{1}{c}{0.29$^{***}$} &
 \multicolumn{1}{c}{0.18$^{***}$} &
 \multicolumn{1}{c}{0.23$^{***}$} & 
 \multicolumn{1}{c}{0.39}\\
 
  & \multicolumn{1}{c}{(0.01)} 
  & \multicolumn{1}{c}{(0.02)} 
  & \multicolumn{1}{c}{(0.003)} 
  & \multicolumn{1}{c}{(0.01)} 
  & \multicolumn{1}{c}{(0.51)}\\ \\
  
\hline \\ [-1.8ex]
Observations & 
\multicolumn{1}{c}{43} &
\multicolumn{1}{c}{109} &
\multicolumn{1}{c}{27} &
\multicolumn{1}{c}{22} &
\multicolumn{1}{c}{5}\\ \\

R$^{2}$ &
\multicolumn{1}{c}{0.88} &
\multicolumn{1}{c}{0.74} &
\multicolumn{1}{c}{0.98} &
\multicolumn{1}{c}{0.94} &
\multicolumn{1}{c}{0.16}\\ \\

Adjusted R$^{2}$ &
\multicolumn{1}{c}{0.88} &
\multicolumn{1}{c}{0.74} &
\multicolumn{1}{c}{0.99} &
\multicolumn{1}{c}{0.95} & 
\multicolumn{1}{c}{$-$0.12}\\ \\

Residual Std. Error & 
 \multicolumn{1}{c}{\begin{tabular}[c]{@{}c@{}} 0.80 \\ df=41\end{tabular}} &
 \multicolumn{1}{c}{\begin{tabular}[c]{@{}c@{}} 0.93 \\ df=107\end{tabular}} &
 \multicolumn{1}{c}{\begin{tabular}[c]{@{}c@{}} 0.39 \\ df=25\end{tabular}} &
 \multicolumn{1}{c}{\begin{tabular}[c]{@{}c@{}} 0.37 \\ df=20\end{tabular}} &
 \multicolumn{1}{c}{\begin{tabular}[c]{@{}c@{}} 2.09 \\ df=3\end{tabular}} \\ \\ 

F Statistic & 
 \multicolumn{1}{c}{\begin{tabular}[c]{@{}c@{}} 324.97$^{***}$ \\ df=1; 41\end{tabular}} &
 \multicolumn{1}{c}{\begin{tabular}[c]{@{}c@{}} 303.32$^{***}$ \\ df=1; 107\end{tabular}} &
 \multicolumn{1}{c}{\begin{tabular}[c]{@{}c@{}} 2,205.71$^{***}$ \\ df=1; 25\end{tabular}} &
 \multicolumn{1}{c}{\begin{tabular}[c]{@{}c@{}} 366.61$^{***}$ \\ df=1; 20\end{tabular}} &
 \multicolumn{1}{c}{\begin{tabular}[c]{@{}c@{}} 0.58 \\ df=1; 3\end{tabular}} \\ \\
 
$\psi$ & 
\multicolumn{1}{c}{1.3803} &
\multicolumn{1}{c}{1.9498} &
\multicolumn{1}{c}{1.5135} &
\multicolumn{1}{c}{1.6982} &
\multicolumn{1}{c}{2.4547}\\ \\

Yearly Computing\\ Power Increase &
\multicolumn{1}{c}{38.0\%} &
\multicolumn{1}{c}{94.9\%} &
\multicolumn{1}{c}{51.3\%} &
\multicolumn{1}{c}{69.8\%} &
\multicolumn{1}{c}{145.4\%}\\ \\ [-1.8ex]
\hline 
\hline \\ [-1.8ex]
\textit{Note:} & \multicolumn{5}{r}{$^{*}$p$<$0.1; $^{**}$p$<$0.05; $^{***}$p$<$0.01}
\end{tabular} 
}
\end{table}

As this shows, for example for Chess, $\frac{\partial log_{10}{(Computing Power)}}{\partial Year} = 0.14$ (statistically significant at p-value $< 1\%$). Exponentiating shows that computer power usage in chess ($\psi=\frac{\partial Computing Power}{\partial Year}$) is 1.38. That is, the amount of computing used for chess has grown by 38\% per year, on average. In other areas, this effect is even stronger, with Go increasing at 95\%, weather prediction by 51\%, oil exploration by 70\% for BP, and protein folding by 145\%.

\subsection{Contributions of I.T. to Performance Improvement}

Table \ref{logperf_logcp} shows our estimates for $\gamma$, the coefficient that describes how increased computing power changes performance across these domains. Here we explicitly model the production function $Y = AL^\alpha K^\beta IT^\gamma$ by taking logs to get our estimating equation: $log{(Y)} = log{(A)}+\alpha\ log{(L)}+\beta\ log{(K)}+\gamma\ log{(IT)}$ where we collectively estimate the non-IT independent variables as part of our residual.

\begin{table}[H] \centering 
 \caption{Performance improvement as computation grows.}
  \label{logperf_logcp} 
\resizebox{\textwidth}{!}{

\begin{tabular}{@{\extracolsep{5pt}}lD{.}{.}{-4} D{.}{.}{-4} D{.}{.}{-4} D{.}{.}{-4} D{.}{.}{-4}} 
\hline 
\\
 & & 
 \multicolumn{1}{c}{\begin{tabular}[c]{@{}c@{}} Computer Chess\end{tabular}} &
 \multicolumn{1}{c}{\begin{tabular}[c]{@{}c@{}} Computer Go\end{tabular}} & 
 \multicolumn{1}{c}{\begin{tabular}[c]{@{}c@{}} Oil Exploration\\ (BP)\end{tabular}} &
 \multicolumn{1}{c}{\begin{tabular}[c]{@{}c@{}} Protein Folding\\ (CASP)\end{tabular}}\\ \\
 
\cline{3-6} \\

 & & 
 \multicolumn{1}{c}{\begin{tabular}[c]{@{}c@{}} ELO\\ ($log_{10}$)\end{tabular}} &
 \multicolumn{1}{c}{\begin{tabular}[c]{@{}c@{}} ELO\\ ($log_{10}$)\end{tabular}} &
 \multicolumn{1}{c}{\begin{tabular}[c]{@{}c@{}} Drilling Success\\ ($log_{10}$)\end{tabular}} &
 \multicolumn{1}{c}{\begin{tabular}[c]{@{}c@{}} GDT\_\ TS\\ ($log_{10}$)\end{tabular}} \\
 
 & & 
 \multicolumn{1}{c}{(1)} &
 \multicolumn{1}{c}{(2)} &
 \multicolumn{1}{c}{(3)} &
 \multicolumn{1}{c}{(5)}\\ \\
 
\hline \\

 $Constant$ & & 
 \multicolumn{1}{c}{3.07$^{***}$} &
 \multicolumn{1}{c}{3.2$^{***}$} &
 \multicolumn{1}{c}{-0.24$^{***}$} &
 \multicolumn{1}{c}{1.43$^{*}$} \\ 
 
  & & 
  \multicolumn{1}{c}{(0.02)} &
  \multicolumn{1}{c}{(0.01)} &
  \multicolumn{1}{c}{(0.05)} &
  \multicolumn{1}{c}{(0.09)} \\ 
  
 $log_{10}(Computing\ Power)$ & &
 \multicolumn{1}{c}{0.05$^{***}$} &
 \multicolumn{1}{c}{0.05$^{***}$} &
 \multicolumn{1}{c}{0.11$^{***}$} &
 \multicolumn{1}{c}{0.04} \\ 
 
  & & 
  \multicolumn{1}{c}{(0.005)} &
  \multicolumn{1}{c}{(0.01)} &
  \multicolumn{1}{c}{(0.011)} &
  \multicolumn{1}{c}{(0.01)}\\ \\
  
\hline 
\\ [-1.8ex]
Observations & &
\multicolumn{1}{c}{31} &
\multicolumn{1}{c}{104} &
\multicolumn{1}{c}{33} &
\multicolumn{1}{c}{5} \\ \\

R$^{2}$ & &
\multicolumn{1}{c}{0.80} &
\multicolumn{1}{c}{0.63} &
\multicolumn{1}{c}{0.78} &
\multicolumn{1}{c}{0.88} \\ \\

Adjusted R$^{2}$ & &
\multicolumn{1}{c}{0.80} &
\multicolumn{1}{c}{0.63} &
\multicolumn{1}{c}{0.77} &
\multicolumn{1}{c}{0.84} \\ \\

Residual Std. Error & &
 \multicolumn{1}{c}{\begin{tabular}[c]{@{}c@{}} 0.06 \\ df=29\end{tabular}} &
 \multicolumn{1}{c}{\begin{tabular}[c]{@{}c@{}} 0.06 \\ df=102\end{tabular}} &
 \multicolumn{1}{c}{\begin{tabular}[c]{@{}c@{}} 0.13 \\ df=31\end{tabular}} &
 \multicolumn{1}{c}{\begin{tabular}[c]{@{}c@{}} 0.03 \\ df=3\end{tabular}} \\ \\
 
F Statistic & & 

 \multicolumn{1}{c}{\begin{tabular}[c]{@{}c@{}} 119.15$^{***}$ \\ df=1; 29\end{tabular}} &
 \multicolumn{1}{c}{\begin{tabular}[c]{@{}c@{}} 174.85$^{***}$ \\ df=1; 102\end{tabular}} &
 \multicolumn{1}{c}{\begin{tabular}[c]{@{}c@{}} 107.80$^{***}$ \\ df=1; 31\end{tabular}} &
 \multicolumn{1}{c}{\begin{tabular}[c]{@{}c@{}} 21.45 \\ df=1; 3\end{tabular}}\\ \\
 
$\gamma$ & &
\multicolumn{1}{c}{0.05} &
\multicolumn{1}{c}{0.05} &
\multicolumn{1}{c}{0.11} &
\multicolumn{1}{c}{0.04} \\
\\ [-1.8ex]
\hline \\

& \multicolumn{1}{c}{} & \multicolumn{1}{c}{} & \multicolumn{1}{c}{Weather Prediction} & \multicolumn{1}{c}{} & \multicolumn{1}{c}{} \\ \\

\cline{2-6} \\

 & \multicolumn{1}{c}{\begin{tabular}[c]{@{}c@{}} Day 3\\ ($log_{10}$)\end{tabular}} &
 \multicolumn{1}{c}{\begin{tabular}[c]{@{}c@{}} Day 4\\ ($log_{10}$)\end{tabular}} &
 \multicolumn{1}{c}{\begin{tabular}[c]{@{}c@{}} Day 5\\ ($log_{10}$)\end{tabular}} &
 \multicolumn{1}{c}{\begin{tabular}[c]{@{}c@{}} Day 6\\ ($log_{10}$)\end{tabular}} &
 \multicolumn{1}{c}{\begin{tabular}[c]{@{}c@{}} Day 7\\ ($log_{10}$)\end{tabular}} \\ 
 
 & \multicolumn{1}{c}{(3)} &
 \multicolumn{1}{c}{(4)} &
 \multicolumn{1}{c}{(5)} &
 \multicolumn{1}{c}{(6)} &
 \multicolumn{1}{c}{(7)} \\ \\
 \hline \\
 $Constant$ & 
 \multicolumn{1}{c}{0.7000$^{***}$} &
 \multicolumn{1}{c}{0.7430$^{***}$} &
 \multicolumn{1}{c}{0.7752$^{***}$} &
 \multicolumn{1}{c}{0.8386$^{***}$} &
 \multicolumn{1}{c}{0.8433$^{***}$} \\
 
  & \multicolumn{1}{c}{(0.0099)}
  & \multicolumn{1}{c}{(0.0119)}
  & \multicolumn{1}{c}{(0.0092)}
  & \multicolumn{1}{c}{(0.0203)}
  & \multicolumn{1}{c}{(0.0167)} \\ 
  
 $log_{10}(Computing\ Power)$ &
 \multicolumn{1}{c}{-0.0355$^{***}$} &
 \multicolumn{1}{c}{-0.0333$^{***}$} &
 \multicolumn{1}{c}{-0.0283$^{***}$} &
 \multicolumn{1}{c}{-0.0313$^{***}$} &
 \multicolumn{1}{c}{-0.0235$^{***}$} \\ 
 
  & \multicolumn{1}{c}{(0.0023)}
  & \multicolumn{1}{c}{(0.0028)}
  & \multicolumn{1}{c}{(0.0022)}
  & \multicolumn{1}{c}{(0.0044)}
  & \multicolumn{1}{c}{(0.0036)}  \\ \\
  
\hline 
\\ [-1.8ex]
Observations &
\multicolumn{1}{c}{22} &
\multicolumn{1}{c}{22} &
\multicolumn{1}{c}{22} &
\multicolumn{1}{c}{18} &
\multicolumn{1}{c}{18} \\ \\

R$^{2}$ &
\multicolumn{1}{c}{0.92} &
\multicolumn{1}{c}{0.88} &
\multicolumn{1}{c}{0.89} &
\multicolumn{1}{c}{0.76} &
\multicolumn{1}{c}{0.73} \\ \\

Adjusted R$^{2}$ &
\multicolumn{1}{c}{0.92} &
\multicolumn{1}{c}{0.87} &
\multicolumn{1}{c}{0.89} &
\multicolumn{1}{c}{0.75} &
\multicolumn{1}{c}{0.71} \\ \\

Residual Std. Error &
\multicolumn{1}{c}{\begin{tabular}[c]{@{}c@{}} 0.02\\ df=20;\end{tabular}} &
\multicolumn{1}{c}{\begin{tabular}[c]{@{}c@{}} 0.03\\ df=20;\end{tabular}} &
\multicolumn{1}{c}{\begin{tabular}[c]{@{}c@{}} 0.02\\ df=20;\end{tabular}} &
\multicolumn{1}{c}{\begin{tabular}[c]{@{}c@{}} 0.02\\ df=16;\end{tabular}} &
\multicolumn{1}{c}{\begin{tabular}[c]{@{}c@{}} 0.02\\ df=16;\end{tabular}} \\ \\

F Statistic & 
\multicolumn{1}{c}{\begin{tabular}[c]{@{}c@{}} 231.67$^{***}$\\ df=1; 20\end{tabular}} &
\multicolumn{1}{c}{\begin{tabular}[c]{@{}c@{}} 141.07$^{***}$\\ df=1; 20\end{tabular}} &
\multicolumn{1}{c}{\begin{tabular}[c]{@{}c@{}} 169.87$^{***}$\\ df=1; 20\end{tabular}} &
\multicolumn{1}{c}{\begin{tabular}[c]{@{}c@{}} 51.54$^{***}$\\ df=1; 16\end{tabular}} &
\multicolumn{1}{c}{\begin{tabular}[c]{@{}c@{}} 43.17$^{***}$\\ df=1; 16\end{tabular}} \\ \\

$\gamma$ & 
\multicolumn{1}{c}{0.04} &
\multicolumn{1}{c}{0.03} &
\multicolumn{1}{c}{0.03} &
\multicolumn{1}{c}{0.03} &
\multicolumn{1}{c}{0.02}\\ \\ [-1.8ex]
\hline \\ [-1.8ex]
\textit{Note:} & \multicolumn{5}{r}{$^{*}$p$<$0.1; $^{**}$p$<$0.05; $^{***}$p$<$0.01}
 \end{tabular} 
}
\end{table}

These show that computing power has, in general, very low exponent $\gamma$, ranging from 0.02 for weather prediction to 0.11 for oil drilling for BP (recall: because weather prediction is measured by error, those values are -$\gamma$). This is what we would expect. Adding a unit of computation to today’s powerful machines has much less impact on outcomes that did ones when computers were brand new.

By contrast with these estimates for the returns to additional computing power, recent estimates for the impact of physical capital on national GDP growth are 0.25-0.4 \cite{crafts2020growth}. That is, the benefit of an additional unit of traditional capital far exceed those for ICT capital. However, so many more units of ICT capital are provided that it again becomes important. For example, according to the Conference Board, although the share of Non-ICT capital in GDP is much higher than ICT capital (by 27 percentage points), the growth of capital services provided by ICT assets is much higher than physical capital (7.5 times faster growth), such that the contribution of capital services provided by ICT versus non-ICT on GDP growth is almost the same \cite{growth_accounting}. 

These results underscore the importance of computing to the economy, but also the need for exponential increases in computing for these contributions to be meaningful.

\subsection{Analysis of variance}

Figure 7 summarizes the $R^2$ results from these regressions, showing the fraction of the variation explained by computing power (dark) and that from all other factors (light).

\begin{figure}[!htp]
	\centering
	\includegraphics[keepaspectratio=true,scale=.34]{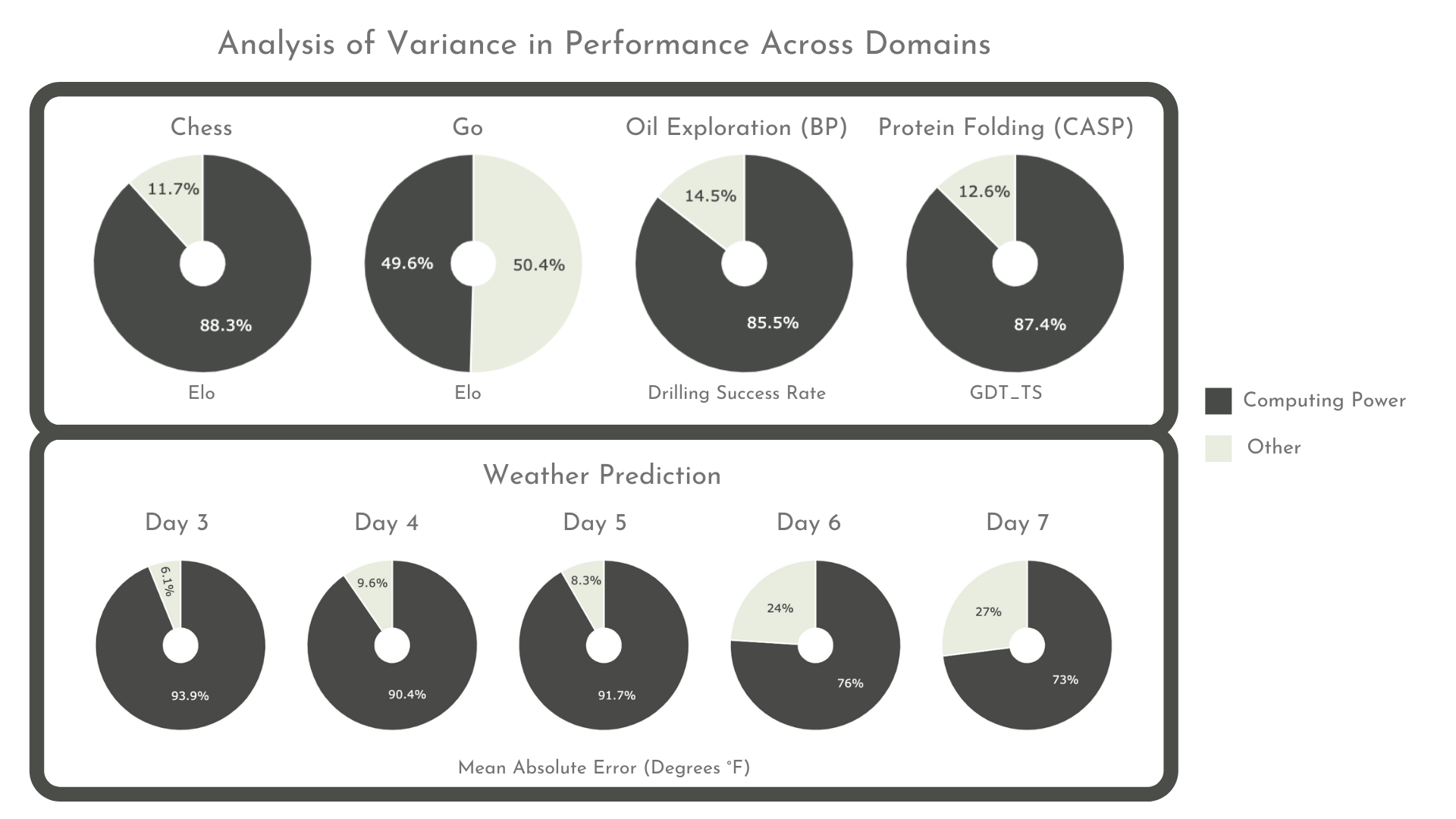}
	\caption{Analysis of Variance in Performance Across Domains.}
	\label{variance_analysis}
\end{figure}

As noted earlier, these results suggest that, in many areas, computing power has been overwhelmingly important as a source of gains in performance.

\subsection{Implications for future performance improvement}

This paper shows that, in the areas of Computer Chess, Computer Go, Weather Prediction, Protein Folding and Oil Exploration, progress depends on getting exponential increases in the amount of computing power. We propose that this is likely a broader phenomenon, because the computational techniques used for these areas are also used in many others. In particular, we suggest that progress in science and engineering often requires finer approximations, more complicated evaluation, repeated analysis under different starting conditions, greater search depth, and more degrees of freedom. Support for this view is provided by Hyperion, a computing research firm, that gathered 690 examples of high-performance computing. They found that investment in High Performance Computing significantly improve economic success and increase scientific innovation across many areas \cite{joseph2013creating}. 

Across our case study areas, Computer Chess, Computer Go, Weather Forecasting, Oil Exploration (BP), and Protein Folding (CASPs), we find that computing power has increased $140\times$, $390{,}000{,}000\times$, $1{,}600{,}000\times$, $160{,}000\times$, $58{,}000\times$ respectively, representing yearly compound growth rates of $14-90\%$.

Fortunately for the procurers of these systems, costs have not risen proportionally to these increases, principally because Moore’s Law provided ever-cheaper computing power \cite{leiserson2020there}. The exact pace of this countervailing improvement in costs from Moore's Law is, however, ambiguous. For example, indexes of this pace are created by macroeconomic agencies, such as the federal reserve banks, and by computer science enthusiasts, for example on Wikipedia \cite{federal_reserve, wiki2021flops}. We find that none of these provide even an approximate fit for the systems where we are able to calculate cost and computing power. For this reason, we estimate the cost index for Computer Chess and Computer Go using actual system costs of programs over time\footnote{Insufficient data was available for the actual costs of weather forecasting systems, protein folding and oil exploration, so we do not include them in this analysis.}. We found that, in Computer Chess, computing power cost has decreased in a rate of 51\% over time. This number is very consistent with what we see in Moore's law ($\sim$48\% per year). However, in Computer Go, costs per gigaflop dropped much more rapidly, droping 117\% per year, approaching the estimates coined by Huang's law \cite{wall2020huang} perhaps because of the usage of GPUs in modern systems.

\begin{figure}[!htp]
	\centering	\includegraphics[keepaspectratio=true,scale=.75]{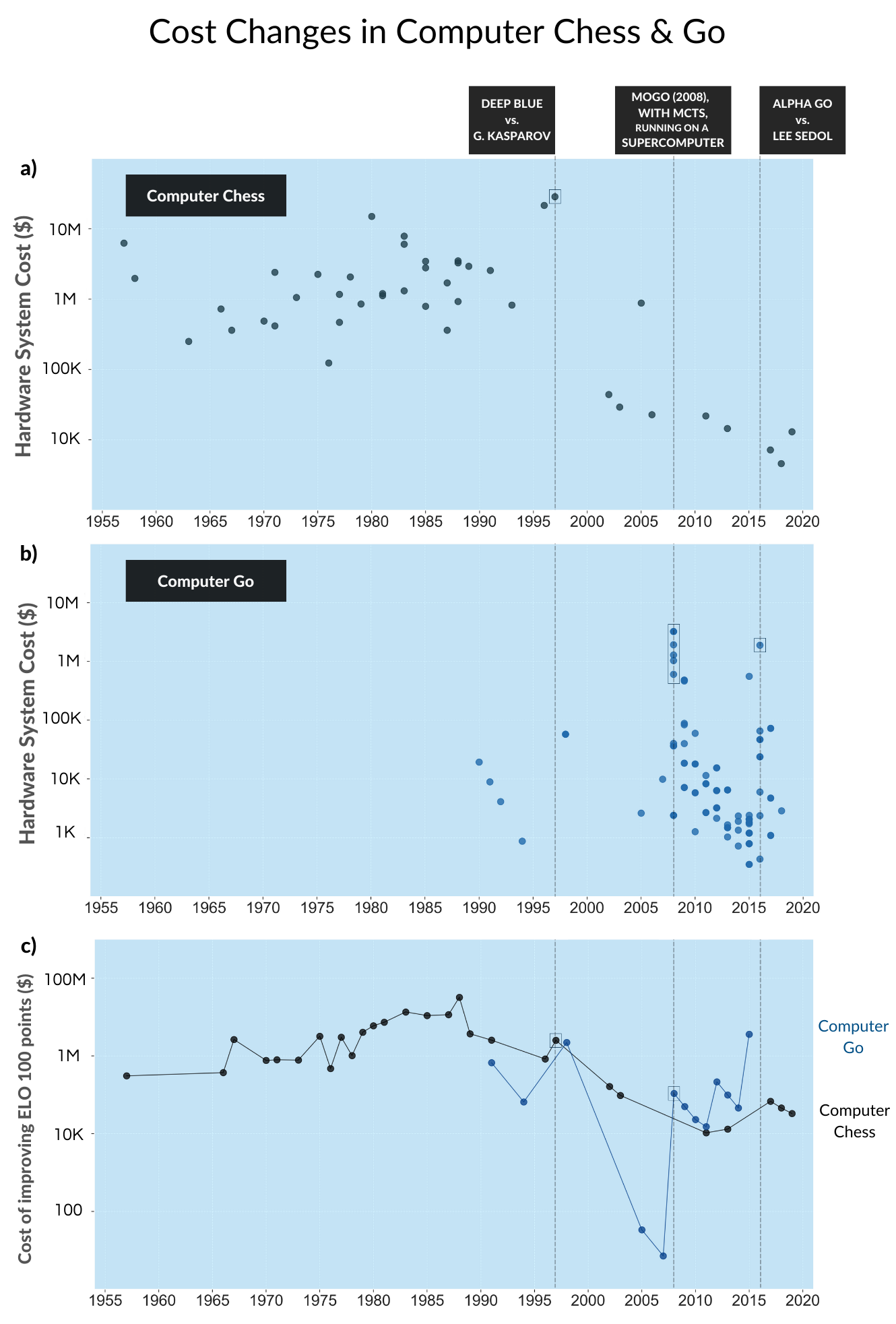}
	\caption{Inferred Cost Changes in Computer Chess \& Go: a) Hardware system cost in Computer Chess. b) Hardware system cost in Computer Go. c) Cost of hardware needed to improve ELO by 100 points in Computer Go and Computer Chess.}
	\label{chess_go_costs}
\end{figure}

Figure \ref{chess_go_costs} show the cost of using ever more computing in these domains, after considering the falling cost of computing power (measured by reported actual system costs across years).

As Figure \ref{chess_go_costs}a shows, the cost of the system used for chess peaked at the time when Deep Blue was playing Kasparov (1996 and 1997) and has since then fallen. This is a good example of a phenomenon identified by Edelman “Unit costs of hardware continue to decline. It is only because usage seems to rise even faster that our total hardware costs continue to rise. It is to be hoped that sooner or later the product of the two - number of operations and cost per operation - will actually decline.” \cite{edelman1981managers}. In Chess, we see that indeed total costs did rise in the lead-up to playing Kasparov, but that post-Kasparov the desire for lower-cost systems won out. 

At the other extreme, the cost of computation in Go (Figure \ref{chess_go_costs}b) had two remarkable moments. The first is mainly due to the emergence of the MCTS algorithm \cite{manso2022mcts}. In a moment of breakthrough in the face of the hope of beating a professional Go player on a 19x19 board, in 2008, the creators of MoGo, in addition to adapting their program with this new algorithm, chose to use computational resources provided by supercomputers \footnote{According to Couetoux, MCTS is an algorithm that generally scales well and it is expected that parallelizing it on a supercomputer would make it perform slightly better \cite{Coutoux2013MonteCT}.}. Eight years later, the cost of computing hit a new peak when AlphaGo defeated Lee Sedol. 

By combining our estimates for the cost of computation and the effect of additional computation on performance, we are also able to show, in Figure c), how expensive it is to increase performance by 100 Elo points over time.
 
\section{Conclusion}

This article provides detailed quantitative case studies of five computing domains: the computing bellwethers of Chess and Go, and the economically important areas of weather prediction, protein folding, and oil exploration. In all cases, we examine the contribution of more computing power to better outcomes, but unlike most previous analyses that analyze based on the spending on IT, we analyze it based on the natural units of computing power. We find that computing power (and implicitly the algorithm changes needed to harness it) account for half or more of all improvement. The size of this contribution is remarkable since a change in computing power has only a tiny effect per unit. But since 1990, the compound growth rates of computing power have ranged from 38\% to 145\% in these areas. With such rapid exponential increases, computing power can nevertheless be the dominant source of improvements in these areas.  Fortunately, such high growth has not come at the full price, since Moore’s Law driven cost decreases made each unit of computing power more affordable. Nevertheless, all of these areas have experienced long stretches of time where costs rose by orders of magnitude.

Overall, this paper paints a coherent picture of computing power improvements as a central driver of progress across many areas over decades, quantifying long-held views about the centrality of I.T. in general, and of Moore’s Law in particular as a driver of long-term performance improvement across society. It shows the importance of exponentially more computing power, and raises the specter of what we would lose absent that growth.

\newpage

%Bibliography
\bibliographystyle{unsrt}  
\bibliography{arxiv.bib}  

\newpage
\section*{Supplemental Materials}\label{supplemental}
\pagenumbering{arabic}

\section{First Differences}\label{appendix:first_diff}

One might worry that correlating two-time trends, both of which are rising over time, might create spurious correlations. We address this by repeating our analysis of how computing power affects performance using a first-differences approach.\footnote{This is also a more strenuous test because it requires immediate causality, whereas we might expect it to take time for software designers and modelers to fully take advantage of new hardware.} 

In each area, we see statistically significant relationships for how computing power affects performance, although for weather we only see if for the Day 3 predictions. 

\begin{table}[H] \centering 
  \label{}
  \caption{First difference analysis}
  
\resizebox{\textwidth}{!}{

\begin{tabular}{@{\extracolsep{5pt}}lccccc}
\\[-1.8ex]\hline
\hline \\
& \multicolumn{5}{c}{\textbf{First Difference}} \\
\cr \cline{2-6} \\
& & Chess
& Go
& \multicolumn{1}{c}{\begin{tabular}[c]{@{}c@{}} Protein Folding\\ (CASP)\end{tabular}}
& \multicolumn{1}{c}{\begin{tabular}[c]{@{}c@{}} Oil Exploration\\ (BP)\end{tabular}} \\
\\ \cline{2-6} \\ & & Elo & Elo & GDT\_TS & Success Rate \\ \\
\hline \\
 $Constant$ & & 24.52$^{}$ & 1.45$^{}$ & -1.84$^{}$ & 0.17$^{}$ \\
  & & (195.14) & (34.79) & (3.56) & (0.10) \\
 $log_{10}(Computing\ Power)$ & & 93.05$^{*}$ & 208.53$^{***}$ & 7.46$^{**}$ & 0.22$^{**}$ \\
  & & (55.26) & (37.09) & (1.22) & (0.10) \\ \\
\hline \\ [-1.8ex]
 Observations & & 82 & 103 & 4 & 27 \\
 $R^2$ & & 0.03 & 0.24 & 0.95 & 0.18 \\
 Adjusted $R^2$ & & 0.02 & 0.24 & 0.92 & 0.15 \\
 Residual Std. Error & & 1766.55(df = 80) & 351.72(df = 101) & 7.12(df = 2) & 0.42(df = 25)  \\
 F Statistic & & 2.83$^{*}$ (df = 1; 80.0) & 31.61$^{***}$ (df = 1; 101) & 37.52$^{**}$ (df = 1; 2) & 5.46$^{**}$ (df = 1; 25) \\
\hline \hline  \\
 & \multicolumn{1}{c}{} & \multicolumn{1}{c}{} & \multicolumn{1}{c}{Weather Prediction} & \multicolumn{1}{c}{} & \multicolumn{1}{c}{} \\ \\
\cline{2-6} \\
 & \multicolumn{1}{c}{Day 3} & \multicolumn{1}{c}{Day 4} & \multicolumn{1}{c}{Day 5} & \multicolumn{1}{c}{Day 6} & \multicolumn{1}{c}{Day 7} \\ 
 & \multicolumn{1}{c}{(3)} & \multicolumn{1}{c}{(4)} & \multicolumn{1}{c}{(5)} & \multicolumn{1}{c}{(6)} & \multicolumn{1}{c}{(7)}\\ \\
\hline \\
 $Constant$ & -0.0337 & -0.0901 & -0.0759 & -0.0586 & -0.0192 \\ 
  & (0.0489) & (0.0745) & (0.0572) & (0.0819) & (0.0855) \\ 
 $log_{10}(Computing\ Power)$ & -0.2127$^{***}$ & -0.0899 & -0.0954 & -0.0598 & -0.1744 \\ 
  & (0.0639) & (0.0973) & (0.0747) & (0.2564) & (0.2676) \\ \\
\hline \\ [-1.8ex]
Observations & \multicolumn{1}{c}{21} & \multicolumn{1}{c}{21} & \multicolumn{1}{c}{21} & \multicolumn{1}{c}{17} & \multicolumn{1}{c}{17} \\ 
R$^{2}$ & \multicolumn{1}{c}{0.3683} & \multicolumn{1}{c}{0.0430} & \multicolumn{1}{c}{0.0790} & \multicolumn{1}{c}{0.0036} & \multicolumn{1}{c}{0.0275} \\ 
Adjusted R$^{2}$ & \multicolumn{1}{c}{0.3350} & \multicolumn{1}{c}{-0.0074} & \multicolumn{1}{c}{0.0306} & \multicolumn{1}{c}{-0.0628} & \multicolumn{1}{c}{-0.0373} \\ 
Residual Std. Error & \multicolumn{1}{c}{$0.1861\ (df = 19)$} & \multicolumn{1}{c}{$0.2834\ (df = 19)$} & \multicolumn{1}{c}{$0.2174\ (df = 19)$} & \multicolumn{1}{c}{$0.2588\ (df = 15)$} & \multicolumn{1}{c}{$0.2702\ (df = 15)$} \\ 
F Statistic & \multicolumn{1}{c}{$11.0769^{***}\ (df = 1; 19)$} & \multicolumn{1}{c}{$0.8536\ (df = 1; 19)$} & \multicolumn{1}{c}{$1.6306\ (df = 1; 19)$} & \multicolumn{1}{c}{$0.0544\ (df = 1; 15)$} & \multicolumn{1}{c}{$0.4247\ (df = 1; 15)$} \\
\hline 
\hline \\
\textit{Note:} & \multicolumn{5}{r}{$^{*}$p$<$0.1; $^{**}$p$<$0.05; $^{***}$p$<$0.01}
\end{tabular}}
\end{table}

\section{Data Sources}\label{appendix:data_sources}

To assess the progress of Computer Chess programs since 1957 (Bernstein’s program), we gathered data from a the resources shown in Table \ref{chess_data}.

\begin{table}[H] \centering 
\label{chess_data}
\caption{Computer Chess Data Sources}
\resizebox{400px}{!}{
\begin{tabular}{|c|l|}
\hline
\textbf{Source}                                            & \textbf{Reference}                                                                                    \\ \hline
Books \& Research Papers
                                                           &  Computer Chess Compendium  \cite{levy2013computer}                                    \\ \cline{2-2} 
                                                           & Computer Games I \cite{levy2012computer}      
                                            \\ \cline{2-2} 
                                                           & Advances in Computer Chess 3 \cite{clarke1982advances}   
                                                                                      \\ \cline{2-2} 
                                                          & Computer Chess: Ten Years of Significant Progress \cite{newborn1989computer}         \\ \cline{2-2} 
                                                           & Chess Skill in Man and Machine \cite{frey2012chess}                                                \\ \cline{2-2} 
                                                           & Concise Encyclopedia of Computer Science \cite{reilly2004concise}                                 \\ \cline{2-2} 
                                                           & Computer Chess, Then And Now: The Deep Blue Saga \cite{hsu1997computer}                            \\ \cline{2-2} 
                                                           & A.C.M monograph series \cite{berliner1976computer}     \\ \cline{2-2} 
                                                           & All About Chess and Computers \cite{levy2012all}                                                     \\ \cline{2-2} 
                                                           & Deep Blue: An Artificial Intelligence Milestone  \cite{newborn2003deep}                         \\ \cline{2-2} 
                                                           & Digital at Work - Snapshots from the first thirty-five years \cite{pearson1992digital}             \\ \cline{2-2} 
                                                           & The Game of Chess \cite{sfetcu2016gaming}                                                                   \\ \cline{2-2} 
                                                           & Kasparov versus Deep Blue \cite{newborn2012kasparov}                                                      \\ \cline{2-2} 
                                                           & Scalable Search in Computer Chess \cite{heinz2013scalable}                                             \\ \cline{2-2} 
                                                           & Beyond Deep Blue Chess in the Stratosphere-Springer-Verlag London \cite{newborn2011}                           \\ \cline{2-2} 
                                                           & The Quest for Artificial Intelligence \cite{nilsson2009quest}                                                       \\ \cline{2-2} 
                                                           & ROBOT, Moravec \cite{moravec2000robot}                                        \\ \cline{2-2} 
                                                           & Rating Computer Science Via Chess \cite{Regan2019}                                      \\ \cline{2-2} 
                                                           & Computers, Chess, and Cognition \cite{schaeffer1990computers}                       \\ \cline{2-2}
                                                           & IBM's Deep Blue Chess Grandmaster Chips \cite{hsu1999ibm}                                                                             \\ \cline{2-2} 
                                                           & The Image Dissector "Eyes" \cite{Horn1969TheID}                                                                           \\ \cline{2-2} 
                                                           & Deep Blue \cite{CAMPBELL200257}                                                                                                   \\ \hline
                                                           Conference and Tournament Reports
                                                                                     & The 22d Annual ACM International Computer Chess Championship \cite{kopec1992the22}    \\ \cline{2-2} 
                                                           &      \begin{tabular}[l]{@{}l@{}} The 23rd ACM International Computer-Chess Championship \\ Report on the Tournament     \cite{Kopec1993The2A}\end{tabular}                                                          \\ \cline{2-2}                           & The 22d Annual ACM International Computer Chess Championship \cite{kopec1992the22}                       \\ \cline{2-2} 
                                                           & Reseña histórica del ajedrez por computadora (V)   \cite{anacadigital1988resena}                                                                \\ \cline{2-2} 
       & Results of ACM’s eighteenth computer chess championship \cite{10.1145/63030.63035}                    \\ \hline
       Newspaper \& Magazine Articles
                                                           & Computerworld Vol.X  \cite{computerworld1976}                                                                                 \\ \cline{2-2} 
                                                           & Computerworld, Vol.XXI \cite{computerworld1987}                                                                  \\ \cline{2-2} 
                                                           & Kasparov flummoxes chess computer   \cite{knight2003kasparov}                     \\ \hline
         Other
                                                           & Computer Chess Rating Lists   \cite{ccrl2021}                                                                        \\ \cline{2-2} 
                                                           & ChessNews \cite{chessnews2021}                                \\  \cline{2-2}
                                                           & Computers and Chess - A History \cite{wall2017computers}                                                          \\ \cline{2-2} 
                                                           & Kaissa \cite{wall2019kaissa}                                                 \\ \cline{2-2} 
                                                           & MacHack Attack \cite{wall2008machack}                                                    \\ \cline{2-2} 
                                                           & KAISSA chess program \cite{maniac2014kaissa}                                                      \\ \cline{2-2} 
                                                           & \begin{tabular}[l]{@{}l@{}} How many positions per second, approximately, was Leela Chess Zero \\ calculating against Tang in the rapid games? \cite{reddit2017how}\end{tabular}  \\ \cline{2-2} 
                                                           & AlphaZero Crushes Stockfish In New 1,000-Game Match  \cite{pete2019alphazero}         \\ \cline{2-2} 
 & Chess Engines and Chess Programs  \cite{wall2021chess}                                              \\ \cline{2-2}
                                                           & Official Website of Chess Genius \cite{chessgenius}                                                                           \\ \cline{2-2} 
                                                           & A Brief History of RISC, the IBM RS/6000 and the IBM eServer pSeries \cite{ibm2021brief}                                               \\ \cline{2-2}                         & Sjeng : a chess-and-variants playing program  \cite{pacutto2021sjeng}                                                                 \\ \hline
\end{tabular}}
\end{table}

To assess the progress of Computer Go Programs, we gathered data from the resources shown in Table \ref{go_data}.

\begin{table}[H] \centering 
\caption{Computer Go Data Sources}
\label{go_data}
\resizebox{450px}{!}{
\begin{tabular}{|c|l|}
\hline
\textbf{Source}                                          & \textbf{Reference}    \\ \hline

Conference \& Tournament Reports        

& Computer Go Challenge at Alternative Party 2009  \cite{raiko2009computer} \\ \cline{2-2}
& Alternative Party 2009 "Man Meets Machine" \cite{altparty2009}         \\ \cline{2-2} 
& 2008 US Go Congress – Computer Go   \cite{computergo2008} \\ \cline{2-2} 
& Cho Chikun 9p defeats AI DeepZen by 2-1   \cite{eurogofed2016cho}   \\ \cline{2-2}
& Interviews with Franz-Josef Dickhut and Rémi Coulom \cite{codecentric2014interviews} \\ \hline

Newspaper \& Magazine Articles          

& Go, going, gone?   \cite{theguardian2009go}  \\ \cline{2-2} 
& Computer Beats Pro at U.S. Go Congress \cite{american2008} \\ \cline{2-2} 
& In Two Moves, AlphaGo and Lee Sedol Redefined the Future \cite{wired2016in} \\ \cline{2-2}
& Microgo1 vs. 3kyu \cite{britgo1984microgo} \\ \hline

Forum Articles \& Community Discussions 

& AlphaGo Zero and the Foom Debate \cite{yudkowsky2017alphago} \\ \cline{2-2} 
& History of Go-playing Programs  \cite{brit2018history}  \\  \cline{2-2}
& \begin{tabular}[l]{@{}l@{}} A Time Line of Supercomputer Go: Temporal \\ Difference Learning to Monte Carlo Programing \cite{usgo2011time}\end{tabular}  \\ \hline

Other

& Computer Go \cite{burmeister1995computer} \\ \cline{2-2}
& A database of go-playing programs \cite{go2018database} \\ \hline

\end{tabular}}
\end{table}

%\section{Weather Prediction}\label{appendix:weather}
\section{Protein Folding}\label{appendix:protein}

Since our main analysis has a limited sample size, we also conduct a second analysis of protein folding’s dependence on computing power increases by analyzing the scaling performance of large distributed folding projects. For instance, the Pande Lab at Stanford University uses 100,000 CPUs distributed throughout the world through the Folding@Home project. Similarly, Rosetta@home, developed in David Baker’s Lab at University of Washington, uses computing from tens of thousands of distributed computers. 
To avail ourselves of all the CASP data, we do this analysis at 3 different levels of protein folding complexity (the levels reflect whether there is a known protein of similar structure that can be used as a guide to decrease the analysis that algorithms must do). These levels range from template-based modeling (TBM), which is easier because of the availability of models, to free modeling (FM) which is harder because it lacks such models. There is also an intermediate level (TBM-FM). 
Using data provided by the Rosetta@home group which performed experiments applying more computing power (by calculating more models for each candidate protein), we observe that for each 10× increase in computation, TBM performance increases by 6\%, TBM/FM performance increases by 8\%, and FM performance improves by 3\% (all statistically significant at p-value $<$ 0.01, see Figure \ref{appendix_protein_table} for details). Their tests show that variance in computing power explain 77-97\% of variance, although this dominance isn’t surprising since other variables aren’t changing simultaneously.

\begin{figure}[!htp]
	\centering
	\includegraphics[keepaspectratio=true,scale=.34]{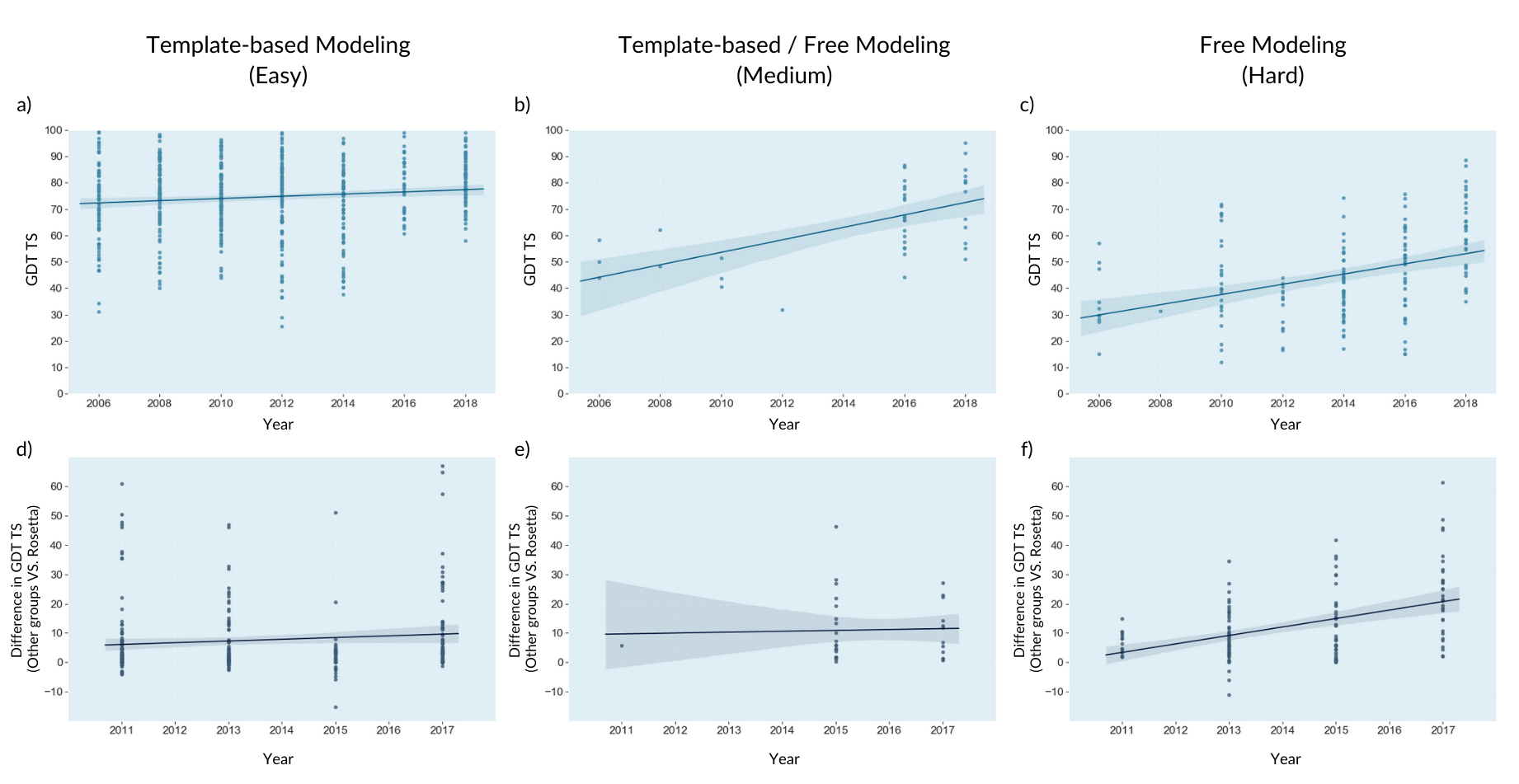}
	\caption{CASP performance in 3 types of protein folding tasks, as measured by protein structure similarity score (GDT-TS) (a-c) Performance (1 dot = 1 model); (d-f) Performance of other server groups compared to to the Rosetta server group that used constant computation.}
	\label{appendix_protein}
\end{figure}

\begin{table}[!htbp] \centering 
  \label{appendix_protein_table}
  \caption{Rosetta@Home: Effect of additional computation on protein folding performance by the difficulty of the folding task.}
\resizebox{.8 \textwidth}{!}{

\begin{tabular}{@{\extracolsep{5pt}}lccc}
\\[-1.8ex]\hline
\hline \\[-1.8ex] 

\\[-1.8ex] & Hard & Medium & Easy \\
\\[-1.8ex] & (1) & (2) & (3) \\ [1.2ex]
\hline \\[-1.8ex] \\
 $Constant$ & 0.321$^{***}$ & 0.656$^{***}$ & 0.770$^{***}$ \\
  & (0.002) & (0.003) & (0.001)  \\ \\
 $log_{10}{(Computing\ Power)}$ & 0.003$^{***}$ & 0.008$^{***}$ & 0.006$^{***}$  \\
  & (0.001) & (0.001) & (0.000) \\ \\
\hline \\[-1.8ex]
 Observations & 10 & 10 & 10 \\
 $R^2$ & 0.771 & 0.844 & 0.969 \\
 Adjusted $R^2$ & 0.742 & 0.824 & 0.965 \\
 Residual Std. Error (df = 8) & 0.002 & 0.003 & 0.001 \\
 F Statistic (df = 1; 8) & 26.892$^{***}$ & 43.110$^{***}$ & 247.901$^{***}$ \\
\hline
\hline \\[-1.8ex]
\textit{Note:} & \multicolumn{3}{r}{$^{*}$p$<$0.1; $^{**}$p$<$0.05; $^{***}$p$<$0.01} \\
\end{tabular}

}
\end{table}

\clearpage

\end{document}